\documentstyle[epsf,graphicx,amstex,12pt]{article}

\input epsf.sty
\begin{document}

\bigskip
\bigskip
\bigskip

{\Huge {\bf Theory of Forces Induced by

 Evanescent Fields}}

\bigskip
\bigskip


{\Large M. Nieto-Vesperinas\footnote{mnieto@@icmm.csic.es} and J. R. Arias-Gonz\'{a}lez\footnote{ricardo.arias@@imdea.org}}

{\it $^1$Instituto de Ciencia de Materiales de Madrid, CSIC}

{\it $^2$Instituto Madrile\~no de Estudios Avanzados en Nanociencia}

{\it Cantoblanco, 28049 Madrid, Spain.}

\bigskip

\section{Introduction}

\label{sec:introduction}

The purpose of this report is to present the theoretical
foundations of the interaction of evanescent fields on an object.
Evanescent electromagnetic waves are inhomogeneous components of the near
field, bound to the surface of the scattering object. These modes travel
along the illuminated sample surface and exponentially decrease outside
it~\cite{bornwolf99ch11,nieto-vesperinas91,mandel95}, {\it e.g.}, either in the
form of lateral waves~\cite{tamir72a,tamir72b} created by total internal
reflection (TIR) at dielectric flat interfaces, whispering--gallery modes in
dielectric tips and particles~\cite
{hill88,owen81,benincasa87,collot93,knight95,weiss95,nieto-vesperinas96} or
of plasmon polaritons~\cite{raether88} in metallic corrugated interfaces
(see Section~\ref{sec:PFM}). The force exerted by these evanescent waves on
particles near the surface is of interest for several reasons. On the one
hand, evanescent waves convey high resolution of the scattered field signal,
beyond the half wavelength limit. This is the essence of near--field
scanning optical microscopy, abbreviated usually as NSOM
\cite{pohl93,paesler96}. These
fields may present large concentrations and intensity enhancements in
subwavelength regions near tips thus giving rise to large gradients that
produce enhanced trapping forces that may enable one to handle particles
within nanometric distances~\cite{novotny97}. In addition, the large
contribution of evanescent waves to the near field is the basis of the high
resolution of signals obtained by transducing the force due to
these waves on particles over surfaces when such particles are used as
probes. On the other hand, evanescent waves have been used both to control
the position of a particle above a surface and to estimate the interaction
(colloidal force) between such a particle and the surface
(see Chapter 6)~\cite{sasaki97,clapp99,dogariu00}.

The first experimental observation, demonstrating the mechanical action of a
single evanescent wave ({\it i.e}., of the lateral wave produced by total
internal reflection at a dielectric (saphire--water interface) on
microspheres immersed in water over a dielectric surface) was made in~\cite
{kawata92}. Further experiments either over waveguides~\cite{kawata96} or
attaching the particle to the cantilever of an atomic force microscope
(AFM)~\cite{vilfan98} aimed at estimating the magnitude of this force.

The scattering of an evanescent electromagnetic wave by a dielectric sphere
has been investigated by several authors using Mie's scattering theory
(addressing scattering cross sections~\cite{chew79} and electromagnetic
forces~\cite{almaas95}), as well as using ray optics~\cite{prieve93,walz99}.
In particular,~\cite{walz99} made a comparison with~\cite{almaas95}.
Although no direct evaluation of either theoretical work with any
experimental result has been carried out yet, likely due to the yet lack
of accurate well characterized and controlled experimental estimations of
these TIR observed forces. In fact, to get an idea of the difficulties of
getting accurate experimental data, we should consider the fluctuations of
the particle position in its liquid environment, due to both Brownian
movement and drift microcurrents, as well as the obliteration produced by
the existence of the friction and van der Waals forces between particle and
surface~\cite{vilfan98,almaas95}. This has led so far to discrepancies
between experiment and theory.

In the next section we shall address the effect of these forces on particles
from the point of view of the dipolar approximation, which is of
considerable interpretative value to understand the contribution of
horizontal and vertical forces. Then we shall show how the multiple
scattering of waves between the surface and the particle introduces
important modifications of the above mentioned forces, both for larger
particles and when they are very close to substrates. Further, we shall
investigate the interplay of these forces when there exists slight
corrugation in the surface profile. Then the contribution of evanescent
waves created under total internal reflection still being important, shares
its effects with radiative propagating components that will exert scattering
repulsive forces. Even so, the particle can be used in these cases as a
scanning probe that transduces this force in a photonic force microscopy
operation.

\section{Force on a Small Particle: \newline
The Dipolar Approximation}

\label{sec:dipapprox}

Small polarizable particles, namely, those with radius $a\ll \lambda $, in
the presence of an electromagnetic field experience a Lorentz
force~\cite{gordon73}:

\begin{equation}  \label{eq:lorentz}
{\bf F}=({\vec {\wp}} \cdot \nabla) {\vec{{\cal E}}}+ \frac{1}{c}\frac{
\partial {\vec \wp}}{\partial t} \times {\vec{{\cal B}}}.
\end{equation}

\noindent
In Equation~(\ref{eq:lorentz}) ${\vec{\wp}}$ is the induced dipole moment
density of the particle, and ${\vec{{\cal E}}}$, ${\vec{{\cal B}}}$ are the
electric and magnetic vectors, respectively.

At optical frequencies, used in most experiments, the observed
magnitude of the electromagnetic force is the time--averaged value. Let the
electromagnetic field be time--harmonic, so that ${\vec{{\cal E}}}({\bf r}
,t)=\Re e \{ {\bf E}({\bf r})\exp (-i\omega t) \}$, ${\vec{{\cal B}}}({\bf r
},t)=\Re e \{ {\bf B}({\bf r})\exp (-i\omega t) \}$, ${\vec{\wp}}({\bf r}
,t)=\Re e \{ {\bf p}({\bf r})\exp (-i\omega t) \}$; ${\bf E}({\bf r})$, $
{\bf B}({\bf r})$ and ${\bf p}({\bf r})$ being complex functions of position
in the space, and $\Re e$ denoting the real part. Then, the time--averaged
Lorentz force over a time interval $T$ large compared to
$2\pi /\omega$~\cite{bornwolf99pp34} is

\begin{equation}
\langle {\bf F}({\bf r})\rangle =\frac{1}{4T}\int_{-T/2}^{T/2}dt\left[ {{{{{(
{\bf p}+{\bf p}^{\ast })\cdot \nabla ({\bf E}+{\bf E}^{\ast })+\frac{1}{c}
\left( \frac{\partial {\bf p}}{\partial t}+\frac{\partial {\bf p}^{\ast }}{
\partial t}\right) \times ({\bf B}+{\bf B}^{\ast })}}}}}\right] ,
\label{eq:averaging}
\end{equation}

\noindent
where $\ast $ denotes complex conjugate. On substituting in
Equation~(\ref{eq:averaging})
${\bf E}$, ${\bf B}$, and ${\bf p}$ by their time harmonic
expressions given above and performing the integral, one obtains for each $
i^{th}$--Cartesian component of the force

\begin{equation}
\langle F_{j}({\bf r})\rangle =\frac{1}{2}\Re e \left\{ p_{k}\frac{\partial
E_{j}^{\ast }({\bf r})}{\partial x_{k}}+\frac{1}{c}\epsilon _{jkl}\frac{
\partial p_{k}}{\partial t}B_{l}^{\ast }\right\} .  \label{eq:chaumetinicio}
\end{equation}

\noindent
In Equation~(\ref{eq:chaumetinicio}) $j=1,2,3$, $\epsilon _{jkl}$ is the
completely antisymmetric Levy--Civita tensor.
On using the Maxwell equation ${\bf B}
=(c/i\omega )\nabla \times {\bf E}$ and the relationships ${\bf p}=\alpha 
{\bf E}$ and $\partial {\bf p}/\partial t=-i\omega {\bf p}$, $\alpha $ being
the particle polarizability, Equation~(\ref{eq:chaumetinicio}) transforms
into

\begin{equation}
\langle F_{j}({\bf r})\rangle =\frac{1}{2}\Re e
\left\{ {{{{{\alpha \left( E_{k}
\frac{\partial E_{j}^{\ast }({\bf r})}{\partial x_{k}}+\epsilon
_{jkl}~\epsilon _{lmn}E_{k}\frac{\partial E_{n}^{\ast}}{\partial x_{m}}
\right) }}}}}\right\} .  \label{eq:chaumetmedio}
\end{equation}

\noindent
Since $\epsilon _{jkl}\epsilon _{lmn}=\delta _{jm}\delta _{kn}-\delta
_{jn}\delta _{km}$, one can finally express the time--averaged Lorentz force
on the small particle as~\cite{chaumet00c}

\begin{equation}
\langle F_{j}({\bf r})\rangle =\frac{1}{2}\Re e \left\{ {{{{{\alpha E_{k}\frac{
\partial E_{k}^{\ast }({\bf r})}{\partial x_{j}}}}}}}\right\} .
\label{eq:chaumetfin}
\end{equation}

\noindent
Equation~(\ref{eq:chaumetfin}) constitutes the expression of the
time--averaged force on a particle in an arbitrary time--harmonic
electromagnetic field.

For a dipolar particle, the polarizability is~\cite{draine88} 

\begin{equation}
\alpha =\frac{\alpha _{0}}{1-\frac{2}{3}ik^{3}\alpha _{0}}.  \label{eq:alfa}
\end{equation}

\noindent
In Equation~(\ref{eq:alfa}) $\alpha _{0}$ is given by: $\alpha
_{0}=a^{3}(\epsilon -1)/(\epsilon +2)$, $\epsilon=\epsilon_2/\epsilon_0$
being the dielectric permittivity contrast between the particle,
$\epsilon_2$, and the surrounding medium, $\epsilon_0$; and
$k=\sqrt{\epsilon_0} k_0$, $k_0=\omega /c$.
For $ka\ll 1$, one can approximate $\alpha $ by:
$\alpha =\alpha _{0}(1+\frac{2}{3}ik^{3}|\alpha _{0}|^{2})$. The imaginary
part in this expression of $\alpha $ constitutes the radiation--reaction term.

The light field can be expressed by its paraxial form, {\it e.g}., it is a
beam or a plane wave, either propagating or evanescent, so that it has a
main propagation direction along ${\bf k}$, the light electric vector will
then be described by

\begin{equation}
{\bf E}({\bf r})={\bf E}_{0}({\bf r})\exp (i{\bf k}\cdot {\bf r}).
\label{eq:oplana}
\end{equation}

\noindent
Substituting Equation~(\ref{eq:oplana}) into Equation~(\ref{eq:chaumetfin}),
one obtains for the force

\begin{equation}
\langle {\bf F}\rangle =\frac{1}{4}\Re e \{ \alpha \} \nabla |{\bf E}
_{0}|^{2}+\frac{1}{2}{\bf k}\Im m \{ \alpha \} |{\bf E}_{0}|^{2}-\frac{1}{2}
\Im m \{ \alpha \} \Im m \{ {\bf E}_{0}\cdot \nabla {\bf E}_{0}^{\ast }\},
\label{eq:finforce}
\end{equation}
where $\Im m$ denotes imaginary part. The first term is the gradient force
acting on the particle, whereas the second term represents the radiation
pressure contribution to the scattering force that, on substituting the
above approximation for $\alpha $, namely, $\alpha =\alpha _{0}(1+\frac{2}{3}
ik^{3}|\alpha _{0}|^{2})$, can also be expressed for a Rayleigh particle
($ka\ll 1$) as~\cite{vandehulst81} $(|{\bf E}_{0}|^{2}/8\pi )C{\bf k}/k$,
where $C$ is the particle scattering cross section given by $C=(8/3)\pi
k^{4}|\alpha _{0}|^{2}$. Notice that the last term of
Equation~(\ref{eq:finforce})
is only zero when either $\alpha $ or ${\bf E}_{0}$ is real.
(This is the case for a plane propagating or evanescent wave but not for a
beam, in general.)

\section{Force on a Dipolar Particle due to \newline
an Evanescent Wave}

\label{sec:dipevanescent}

Let the small particle be exposed to the electromagnetic field of an
evanescent wave, whose electric vector is ${\bf E}={\bf T}\exp (-qz)\exp (i
{\bf K}\cdot {\bf R})$, where we have written ${\bf r}=({\bf R},z)$ and
${\bf k}=({\bf K},k_{z})$, ${\bf K}$ and $k_{z}$ satisfying
$K^{2}+k_{z}^{2}=k^{2}$, $k^{2}=\omega ^{2}\epsilon _{0}/c^{2}$,
with $k_{z}=iq=i\sqrt{K^{2}-k_{0}^{2}}$. This field is created under total
internal reflection at a flat interface ($z=$ constant, below the particle)
between two media of dielectric permittivity ratio $\epsilon_0/\epsilon_1$
(see also
inset of Figure~\ref{fig:dielec}(a)). The incident wave, either $s$ or $p$
polarized, ({\it i.e}., with the electric vector either perpendicular or in
the plane of incidence: the plane formed by the incident wavevector ${\bf k}
_{i}$ at the interface and the surface normal $\hat{z}$) enters from the
denser medium at $z<0$. The particle is in the medium at $z>0$. Without loss
of generality, we shall choose the incidence plane as $OXZ$, so that ${\bf K}
=(K,0)$. Let $T_{\perp }$ and $T_{\parallel }$ be the transmitted amplitudes
into $z>0$ for $s$ and $p$ polarizations, respectively. The electric vector
is:

\begin{equation}
{\bf E}=(0,1,0){T_{\perp }}\exp (iKx)\exp (-qz),  \label{eq:evanescent1}
\end{equation}

\noindent
for $s$ polarization, and

\begin{equation}
{\bf E}=(-iq,0,K)\frac{T_{\parallel }}{k}\exp (iKx)\exp (-qz).
\label{eq:evanescent2}
\end{equation}

\noindent
for $p$ polarization

By introducing the above expressions for the electric vector ${\bf E}$ into
Equation~(\ref{eq:finforce}), we readily obtain the average total force on
the particle split into the scattering and gradient forces. The scattering
force is contained in the $OXY$--plane (that is, the plane containing the
propagation wavevector of the evanescent wave), namely,

\begin{equation}  \label{eq:evanescent3}
\langle F_x \rangle = \frac{|T|^2}{2} K \Im m \{ \alpha \} \exp(-2qz);
\end{equation}

\noindent
For the gradient force, which is purely directed along $OZ$, one has

\begin{equation}  \label{eq:evanescent4}
\langle F_z \rangle = -\frac{|T|^2}{2}q \Re e \{ \alpha \} \exp(-2qz).
\end{equation}

\noindent
In Equations~(\ref{eq:evanescent3}) and~(\ref{eq:evanescent4}) $T$ stands
for either $T_{\perp }$ or $T_{\parallel }$, depending on whether the
polarization is $s$ or $p$, respectively.

For an absorbing particle, on introducing Equation~(\ref{eq:alfa}) for
$\alpha$ into Equations~(\ref{eq:evanescent3}) and~(\ref{eq:evanescent4}),
one gets for the scattering force

\begin{equation}  \label{eq:fx1}
\langle F_x \rangle = \frac{|T|^2}{2} K \exp(-2qz)\frac{\Im m \{ \alpha_0 \}
+(2/3)k^3 |\alpha_0|^2}{1+(4/9)k^6|\alpha_0|^2},
\end{equation}

\noindent
and for the gradient force

\begin{equation}  \label{eq:fz1}
\langle F_z \rangle = -\frac{|T|^2}{2}q \frac{\Re e \{ \alpha_0 \} }{1+(4/9)k^6|
\alpha_0|^2} \exp(-2qz).
\end{equation}

It should be remarked that, except near resonances, in general
$\Im m \{ \alpha _{0} \}$ is a positive quantity
and therefore the scattering force in Equation~(\ref{eq:fx1}) is positive in
the propagation direction $K$ of the evanescent wave, thus pushing the
particle parallel to the surface, whereas the gradient force
Equation~(\ref{eq:fz1})
is negative or positive along $OZ$, therefore, attracting or
repelling the particle towards the surface, respectively, according to whether
$\Re e \{ \alpha \} > 0$ or $\Re e \{ \alpha \} < 0$.
The magnitudes of these forces increase with the decrease of
distance to the interface and it is larger for $p$ polarization since in
this case the dipoles induced by the electric vector at both the particle
and the surface are oriented parallel to each other, thus resulting in a
smaller interaction than when these dipoles are induced in the $OXZ$--plane
($s$--polarization)~\cite{alonsofinn68}.

In particular, if $ka\ll 1$, Equation~(\ref{eq:fx1}) becomes

\begin{equation}  \label{eq:fx2}
\langle F_x \rangle = \frac{|T|^2}{2} K \exp(-2qz) \left[a^3 \Im m \left\{
\frac{\epsilon-1} {\epsilon+2}\right \} + \frac{2}{3}k^3 a^6 \left |\frac{
\epsilon-1}{\epsilon+2} \right|^2 \right];
\end{equation}

\noindent
The first term of Equation~(\ref{eq:fx2}) is the radiation pressure of the
evanescent wave on the particle due to absorption, whereas the second term
corresponds to scattering. This expression can be further expressed as

\begin{equation}  \label{eq:fx3}
\langle F_x \rangle = \frac{|T|^2}{8 \pi} \frac{K}{k} \exp(-2qz)~C_{ext}.
\end{equation}

\noindent
where the particle extinction cross section $C_{ext}$ has been introduced as

\begin{equation}  \label{eq:eficaz}
C_{ext}=4\pi k a^3 \Im m \left \{ \frac{\epsilon-1}{\epsilon+2} \right\}
+\frac{8
\pi}{3}k^4 a^6 \left |\frac{\epsilon-1}{\epsilon+2} \right|^2.
\end{equation}

\noindent
Notice that Equation~(\ref{eq:eficaz}) coincides with the value obtained
from Mie's theory for small particles in the low--order expansion of the size
parameter $ka$ of the extinction cross section~\cite{vandehulst81}.

Although the above equations do not account either for multiple scattering
as described by Mie's theory for larger particles or for multiple interactions
of the wave between the particle and the dielectric surface, they are useful
to understand the fundametals of the force effects induced by a single
evanescent wave on a particle. It should be remarked, however, that as shown
at the end of this section, once Mie's theory becomes necessary,
multiple scattering with the surface demands that its contribution be taken
into account.

\begin{figure}[h]
\begin{center}
\includegraphics*[draft=false,width=\linewidth]{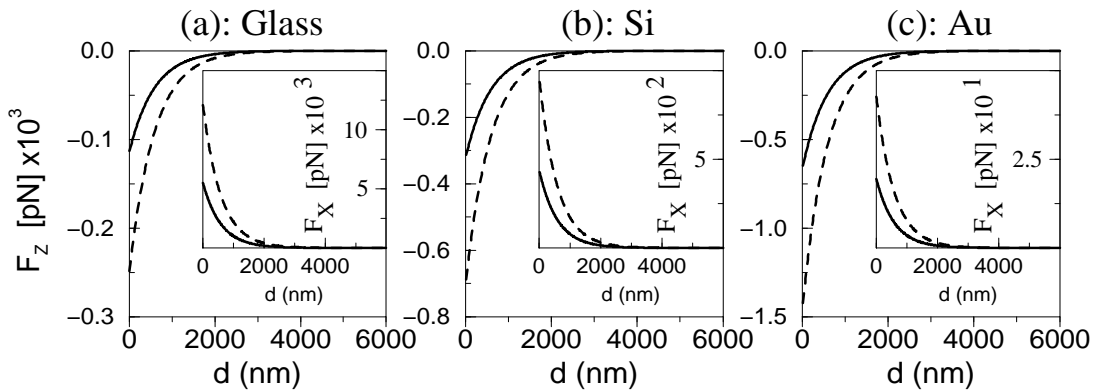}
\end{center}
\caption{ Forces in the $Z$ direction and in the $X$ direction (as insets)
acting on a sphere with radius $a=60$ $nm$, in the dipolar approximation.
The angle of incidence is $\protect\theta _0=42^o$, larger than the critical
angle $\protect\theta _c=41.8^o$ (for the glass--air interface). $\protect
\lambda=632.8$ $nm$. Solid lines: $s$ polarization, dashed lines: $p$
polarization. The sphere material is: (a): glass, (b): silicon, (c): gold.}
\label{fig:dipolo}
\end{figure}

Figure~\ref{fig:dipolo} shows the evolution of the scattering and gradient
forces on three kinds of particles, namely, glass ($\epsilon_2 =2.25$),
silicon ($\epsilon_2 =15+i0.14$) and gold ($\epsilon_2 =-5.65+i0.75$), all of
radius $a=60$ $nm,$ as functions of the gap distance $d$ between the
particle and the surface at which the evanescent wave is created. The
illuminating evanescent wave is due to refraction of an incident plane wave
of power $P=\frac{c\sqrt{\epsilon_1 }}{8\pi }|A|^{2}=1.9\times 10^{-2}$
$mW/\mu m^{2}$, equivalent to $150$ $mW$ over a circular section of radius
$50$ $\mu m$, on a glass--air interface at angle of incidence $\theta
_{0}=42^{o} $ and $\lambda =632.8$ $nm$ (the critical angle is $\theta
_{c}=41.8^{o}$), both at $s$ and $p$ polarization (electric vector
perpendicular and parallel, respectively, to the incidence plane at the
glass--air interface, namely, $|T_{\perp }|^{2}=4\epsilon_1 \cos ^{2}\theta
_{0}|A|^{2}/(\epsilon_1 -1) $, $|T_{\parallel }|^{2}=4\epsilon_1 \cos ^{2}\theta
_{0}|A|^{2}/[(\epsilon_1 -1)((1+\epsilon_1 )\sin ^{2}\theta _{0}-1)]$).

These values of forces are consistent with the magnitudes obtained on
similar particles by applying Maxwell's stress tensor (to be discussed in
Section~\ref{sec:CDM}) via Mie's scattering theory~\cite{almaas95}. However,
as shown in the next section, as the size of the particle increases, the
multiple interaction of the illuminating wave between the particle and the
substrate cannot be neglected. Therefore, the above results, although of
interpretative value, should be taken with care at distances smaller than
$10$ $nm$, since in that case multiple scattering makes the force stronger.
This will be seen next.

\section{Influence of Interaction with the Substrate}

\label{sec:substrate}

Among the several studies on forces of evanescent waves over particles,
there are several models which calculate the forces from Maxwell's stress
tensor on using Mie's theory to determine the scattered field in and around
the sphere, without, however, taking into account the multiple
scattering between the interface at which the evanescent wave is created and
the sphere~\cite{almaas95,chang94,walz99}. We shall next see that, except
at certain distances, this multiple interaction cannot be neglected.

The equations satisfied by the electric and magnetic vectors in a
non--magnetic medium are

\begin{eqnarray}
\nabla \times \nabla \times {\bf E}-k^{2}{\bf E} & = & 4\pi k^{2}{\bf P},
\label{eq:green1}
\\ [+3mm]
\nabla \times \nabla \times {\bf H}-k^{2}{\bf H} & = &
-i4\pi k\nabla \times {\bf P},
\label{eq:green2}
\end{eqnarray}

\noindent
where ${\bf P}$ is the polarization vector. The solutions to
Equations~(\ref{eq:green1}) and~(\ref{eq:green2})
are written in integral form as

\begin{eqnarray}
{\bf E}({\bf r}) & = & k^{2}\int d^{3}r^{\prime }~{\bf P}({\bf r}^{\prime }) 
\cdot\overset{\leftrightarrow }{{\cal G}}({\bf r},{\bf r}^{\prime }),
\label{eq:green3}
\\ [+3mm]
{\bf H}({\bf r}) & = & -ik\int d^{3}r^{\prime }~\nabla \times {\bf P}({\bf r}
^{\prime })\cdot \overset{\leftrightarrow }{{\cal G}}({\bf r},{\bf r}
^{\prime }),
\label{eq:green4}
\end{eqnarray}

\noindent
In Equations~(\ref{eq:green3}) and~(\ref{eq:green4}) $\overset{
\leftrightarrow }{{\cal G}}({\bf r},{\bf r}^{\prime })$ is the outgoing
Green's dyadic or field created at ${\bf r}$ by a point dipole at ${\bf r}
^{\prime }$. It satisfies the equation

\begin{equation}
\nabla \times \nabla \times \overset{\leftrightarrow }{{\cal G}}({\bf r},
{\bf r}^{\prime })-k^{2}\overset{\leftrightarrow }{{\cal G}}({\bf r},{\bf r}
^{\prime })=4\pi \delta ({\bf r}-{\bf r}^{\prime })\overset{\leftrightarrow }
{{\cal I}}.  \label{eq:green5}
\end{equation}

Let's introduce the electric
displacement vector ${\bf D} = {\bf E} + 4 \pi {\bf P}$. Then,

\begin{equation}
\nabla \times \nabla \times {\bf E} = \nabla \times \nabla \times {\bf D} -
4 \pi \nabla \times \nabla \times {\bf P}.  \label{eq:identity1}
\end{equation}

\noindent
Using the vectorial identity $\nabla \times \nabla \times {\bf D} =
\nabla (\nabla \cdot {\bf D}) - \nabla ^2 {\bf D}$, and the fact that in the
absence of free charges, $\nabla \cdot {\bf D}=0$, it is easy to obtain

\begin{eqnarray}
\nabla \times \nabla \times {\bf E} & = & - \nabla ^2 {\bf D} - 4 \pi
\left[ \nabla (\nabla \cdot {\bf P}) - \nabla ^2 {\bf P} \right] \\
[+3mm]
& = &
- \nabla ^2 {\bf E} - 4 \pi \nabla (\nabla \cdot {\bf P}),
\label{eq:identity2}
\end{eqnarray}

\noindent 
which straightforwardly transforms Equation~(\ref{eq:green1}) into:

\begin{equation}
\nabla ^{2}{\bf E}+k^{2}{\bf E}=-4\pi \lbrack k^{2}{\bf P}+\nabla (\nabla
\cdot {\bf P})],  \label{eq:green6}
\end{equation}

\noindent
whose solution is

\begin{equation}
{\bf E}({\bf r})=k^{2}\int d^{3}r^{\prime }~[{\bf P}+\nabla (\nabla \cdot 
{\bf P})]({\bf r}^{\prime })~G({\bf r},{\bf r}^{\prime }) \, .
\label{eq:electrico}
\end{equation}

\noindent
In a homogenous infinite space the function $G$ of
Equation~(\ref{eq:electrico})
is $G_{0}({\bf r},{\bf r}^{\prime })=\exp (ik|{\bf r}-{\bf r}
^{\prime }|)/|{\bf r}-{\bf r}^{\prime }|$, namely, a spherical wave, or
scalar Green's function, corresponding to radiation from a point source at
${\bf r}^{\prime }$.

To determine $\overset{\leftrightarrow }{{\cal G}}$, we consider the case in
which the radiation comes from a dipole of moment ${\bf p}$, situated at
${\bf r}_{0}$; the polarization vector ${\bf P}$ is expressed as

\begin{equation}
{\bf P}({\bf r})={\bf p}\delta ({\bf r}-{\bf r}_{0}).
\label{eq:polarizacion}
\end{equation}

\noindent
Introducing Equation~(\ref{eq:polarizacion}) into
Equation~(\ref{eq:electrico})
one obtains the well known expression for the electric field
radiated by a dipole

\begin{equation}
{\bf E}({\bf r})=k^{2}{\bf p}\nabla ({\bf p}\cdot \nabla )\frac{\exp (ik|
{\bf r}-{\bf r}^{\prime }|)}{|{\bf r}-{\bf r}^{\prime }|}.
\label{eq:dipole1}
\end{equation}

\noindent
On the other hand, if Equation~(\ref{eq:polarizacion}) is introduced into
Equation~(\ref{eq:green3}) one obtains

\begin{equation}
{\bf E}({\bf r})=k^{2}{\bf p}\cdot \overset{\leftrightarrow }{{\cal G}}_{0}(
{\bf r},{\bf r}^{\prime }).  \label{eq:dipole2}
\end{equation}

\noindent
On comparing Equations~(\ref{eq:dipole1}) and~(\ref{eq:dipole2}), since both
give identical value for ${\bf E}$, we get

\begin{equation}
k^{2}{\bf P}({\bf r}^{\prime })\cdot \overset{\leftrightarrow }{{\cal G}}
_{0}({\bf r},{\bf r}^{\prime })=[k^{2}{\bf p}\nabla ({\bf p}\cdot \nabla
)]G_{0}({\bf r},{\bf r}^{\prime }),  \label{eq:dyadic1}
\end{equation}

\noindent
{\it i.e}., the tensor Green's function in a homogenous infinite space is

\begin{equation}
\overset{\leftrightarrow }{{\cal G}}_{0}({\bf r},{\bf r}^{\prime })=\left( 
\overset{\leftrightarrow }{{\cal I}}+\frac{1}{k^{2}}\nabla \nabla \right)
G_{0}({\bf r},{\bf r}^{\prime }).  \label{eq:dyadic2}
\end{equation}

A remark is in order here. When applying Equation~(\ref{eq:dyadic2}) in
calculations one must take into account the singularity at ${\bf r}={\bf r}
^{\prime }$, this is accounted for by writing $\overset{\leftrightarrow }{
{\cal G}}_{0}$ as~\cite{yaghjian80}:

\begin{equation}
\overset{\leftrightarrow }{{\cal G}}_{0}({\bf r},{\bf r}^{\prime })={\cal P}
\left[ {{{{{{{{\ \left( \overset{\leftrightarrow }{{\cal I}}+\frac{1}{k^{2}}
\nabla \nabla \right) G_{0}({\bf r},{\bf r}^{\prime })}}}}}}}}\right] -\frac{
1}{k^{2}}\delta ({\bf r}-{\bf r}^{\prime })\overset{\leftrightarrow }{{\bf L}
}_{v}.  \label{eq:dyadic3}
\end{equation}

\noindent
In Equation~(\ref{eq:dyadic3}) ${\cal P}$ represents the principal value and 
$\overset{\leftrightarrow }{{\bf L}}_{v}$ is a dyadic that describes the
singularity and corresponds to an exclusion volume around ${\bf r}={\bf r}
^{\prime }$, on whose shape it depends~\cite{yaghjian80}.

\subsection{The Coupled Dipole Method}

\label{sec:CDM}

Among the several methods of calculating multiple scattering between bodies
of arbitrary shape ({\it e.g}., transition matrix, finite--difference time
domain, integral procedures, discrete dipole approximation, {\it etc}.) we
shall next address the {\em coupled dipole method} (Purcell and
Pennipacker~\cite{purcell73}).
This procedure is specially suitable for multiple
scattering between a sphere and a flat interface.

Let us return to the problem of determining the interaction of the incident
wave with the substrate and the sphere. The scattered electromagnetic field
is obtained from the contribution of all polarizable elements of the system
under the action of the illuminating wave. The electric vector above the
interface is given by the sum of the incident field ${\bf E}_{i}$ and that
expressed by Equation~(\ref{eq:green3}) with the dyadic Green function
$\overset{\leftrightarrow }{{\cal G}}$ being given by

\begin{equation}
\overset{\leftrightarrow }{{\cal G}}({\bf r},{\bf r}^{\prime })=\overset{
\leftrightarrow }{{\cal G}}_{0}({\bf r},{\bf r}^{\prime })+\overset{
\leftrightarrow }{{\cal G}}_{s}({\bf r},{\bf r}^{\prime }).
\label{eq:sumdyadic}
\end{equation}

\noindent
In Equation~(\ref{eq:sumdyadic}) $\overset{\leftrightarrow }{{\cal G}}_{0}$
is given by Equation~(\ref{eq:dyadic2}) and, as such, it corresponds to the
field created by a dipole in a homogeneous infinite space. On the other
hand, $\overset{\leftrightarrow }{{\cal G}}_{s}$ represents the field from
the dipole after reflection at the interface.

\begin{figure}[t]
\begin{center}
\includegraphics*[draft=false,width=\linewidth]{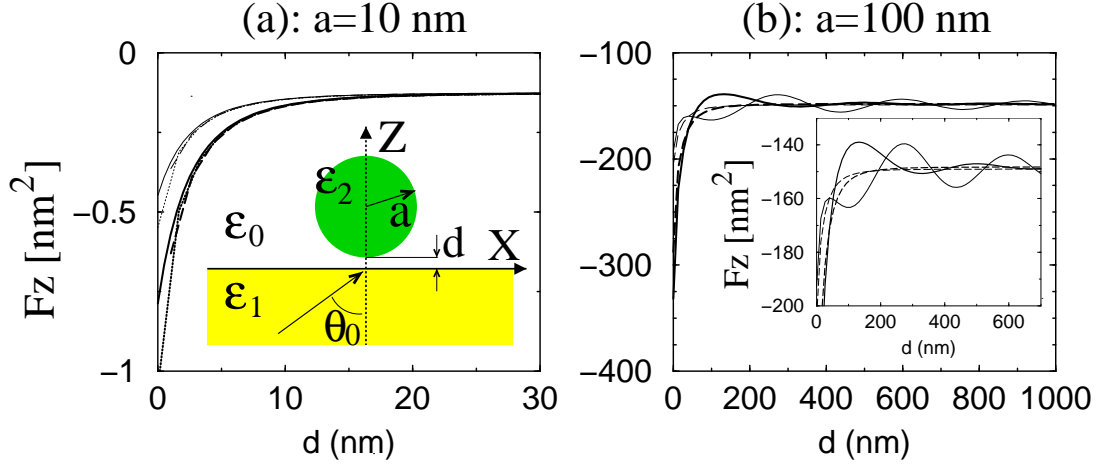}
\end{center}
\caption{ Normalized force in the $Z$ direction acting on a glass sphere on
a glass--vacuum interface. The angle of incidence $\protect\theta_ 0=42 ^o$
is larger than the critical angle $\protect\theta_ c=41.8 ^o$. $\protect
\lambda=632.8$ $nm$. Thin lines: $S$ polarization, Thick lines: $P$
polarization. (a): $a=10$ $nm$, full line: dipole approximation, dashed
line: CDM--A, dotted line: CDM--B. The inset shows the scattering geometry.
(b): $a=100$ $nm$, full line: calculation with CDM--B, dashed line: static
approximation. (From Ref.~\protect\cite{chaumet00a}). }
\label{fig:dielec}
\end{figure}

The polarization vector ${\bf P}$ is represented by the collection of $N$
dipole moments ${\bf p}_j$ corresponding to the $N$ polarizable elements of
all materials included in the illuminated system, namely,

\begin{equation}  \label{eq:sumpolarizacion}
{\bf P}({\bf r})=\sum_{j}^{N}{\bf p}_j\delta({\bf r}-{\bf r}_i).
\end{equation}

\noindent
The relationship between the $k^{{\rm th}}$ dipole moment ${\bf p}_{k}$ and
the exciting electric field is, as before, given by ${\bf p}_{k}=\alpha _{k}
{\bf E}({\bf r}_{k})$, with $\alpha _{k}$ expressed by
Equation~(\ref{eq:alfa}).
Then, Equations~(\ref{eq:electrico}),~(\ref{eq:sumdyadic})
and~(\ref{eq:sumpolarizacion}) yield

\begin{equation}
{\bf E}({\bf r}_{j})=k^{2}\sum_{k}^{N}\alpha _{k}[
\overset{\leftrightarrow }{{\cal G}}_{0}({\bf r}_{j},{\bf r}_{k})+\overset{
\leftrightarrow }{{\cal G}}_{s}({\bf r}_{j},{\bf r}_{k})]\cdot {\bf E}({\bf r
}_{k}).  \label{eq:electricdip}
\end{equation}

The determination of $\overset{\leftrightarrow }{{\cal G}}_{s}$ either above
or below the flat interface is discussed next (one can find more details in
Ref.~\cite{agarwal75a}). Let us summarize the derivation of its expression
above the surface. The field ${\bf E}$ in the half--space $z>0$, from a
dipole situated in this region, is the sum of that from the dipole in free
space and the field ${\bf E}_{r}$ produced on reflection of the latter at
the interface. Taking Equation~(\ref{eq:dipole2}) into account, this is
therefore

\begin{equation}
{\bf E}({\bf r})=k^{2}{\bf p}\cdot \overset{\leftrightarrow }{{\cal G}}_{0}(
{\bf r},{\bf r}^{\prime })+{\bf E}_{r}({\bf r}),\text{ \ \ \ }z>0 \, .
\label{eq:electricdipfin}
\end{equation}

\noindent
Both the spherical wave $G_{0}$ and ${\bf E}_{r}$ are expanded into plane
waves. The former is, according to Weyl's
representation~\cite{banios66ch,nieto-vesperinas91},

\begin{equation}
G_{0}({\bf r},{\bf r}^{\prime })=\frac{i}{2\pi }\int_{-\infty }^{\infty }
\frac{d^{2}K}{k_z ({\bf K})}\exp [i({\bf K}\cdot ({\bf R}-{\bf R}^{\prime
})+k_z |z-z^{\prime }|)] \, .  \label{eq:weyl}
\end{equation}

\noindent
On the other hand, ${\bf E}_{r}$ is expanded as an angular spectrum of plane
waves \cite{nieto-vesperinas91,mandel95}

\begin{equation}
{\bf E}_{r}({\bf r})=\int_{-\infty }^{\infty }d^{2}K~{\bf A}_{r}({\bf K}
)\exp [i({\bf K}\cdot {\bf R}+k_z z)].  \label{eq:angularspectrum}
\end{equation}

\noindent
Introducing Equation~(\ref{eq:weyl}) into~Equation (\ref{eq:sumdyadic}) one
obtains a plane wave expansion for $\overset{\leftrightarrow }{{\cal G}}_{0}$.
This gives the plane wave components ${\bf A}_{h}({\bf K})$ of the first
term of Equation~(\ref{eq:electricdipfin}). Then, the plane wave components
${\bf A}_{r}({\bf K})$ of the second term of Equation~(\ref{eq:electricdipfin})
are given by

\begin{equation}
{\bf A}_{r}({\bf K})=r({\bf K}){\bf A}_{h}({\bf K}),  \label{eq:amplitude}
\end{equation}

\noindent
In Equation~(\ref{eq:amplitude}) $r({\bf K})$ is the Fresnel reflection
coefficient corresponding to the polarization of ${\bf A}_{h}$. The result
is therefore that $\overset{\leftrightarrow }{{\cal G}}$,
Equation~(\ref{eq:sumdyadic}), is

\begin{equation}
\overset{\leftrightarrow }{{\cal G}}({\bf r},{\bf r}^{\prime })=\frac{1}{
4\pi ^{2}}\int_{-\infty }^{\infty }d^{2}K~{\overset{\leftrightarrow }{{\bf S}
}}^{-1}({\bf K})\cdot \overset{\leftrightarrow }{{\bf g}}({\bf K}
,z,z^{\prime })\cdot \overset{\leftrightarrow }{{\bf S}}({\bf K})\exp [i{\bf 
K}\cdot ({\bf R}-{\bf R}^{\prime })],  \label{eq:CDMdyadic}
\end{equation}

\noindent
where~\cite{agarwal75a,agarwal75b,keller93}

\begin{equation}
\overset{\leftrightarrow }{{\bf S}}({\bf K})=\frac{1}{K}
\left( \begin{array}{ccc}
k_{x} & k_{y} & 0 \\ 
-k_{y} & k_{x} & 0 \\ 
0 & 0 & K
\end{array}
\right) ,  \label{eq:Stensor}
\end{equation}

\noindent
and the dyadic $\overset{\leftrightarrow }{{\bf g}}$ has the
elements~\cite{greffet97}

\begin{eqnarray}
g_{11} & = & \frac{-ik_z^{(0)}}{2\epsilon _{0}k_0^{2}}\left[ \frac{\epsilon
_{0}k_z^{(1)}-\epsilon _{1}k_z^{(0)}}
{\epsilon _{0}k_z^{(1)}+\epsilon _{1}k_z^{(0)}}\exp
[ik_z^{(0)}(z+z^{\prime })]+\exp (ik_z^{(0)}|z-z^{\prime }|)\right] ,
\label{eq:g11}
\\ [+3mm]
g_{22} & = & \frac{-i}{2k_z^{(0)}}\left[ {{{{{\frac{k_z^{(0)}-k_z^{(1)}}
{k_z^{(0)}+k_z^{(1)}}\exp
[ik_z^{(0)}(z+z^{\prime })]+\exp (ik_z^{(0)}|z-z^{\prime }|)}}}}}\right] ,
\label{eq:g22}
\\ [+3mm]
g_{33} & = & \frac{iK^{2}}{2\epsilon _{0}k_z^{(0)}k_0^{2}}\left[ \frac{\epsilon
_{0}k_z^{(1)}-\epsilon _{1}k_z^{(0)}}{\epsilon _{0}k_z^{(1)}+
\epsilon _{1}k_z^{(0)}}\exp
[ik_z^{(0)}(z+z^{\prime })]-\exp (ik_z^{(0)}|z-z^{\prime }|)\right]
\nonumber \\ [+3mm]
& + &
\frac{1}{\epsilon _{0}k_0^{2}}\delta (z-z^{\prime }),
\label{eq:g33}
\\ [+3mm]
g_{12} & = & 0,  \label{eq:g12}
\\ [+3mm]
g_{13} & = & \frac{-iK}{2\epsilon _{0}k_0^{2}}\left[ \frac{\epsilon
_{0}k_z^{(1)}-\epsilon _{1}k_z^{(0)}}{\epsilon _{0}k_z^{(1)}+\epsilon _{1}k_z^{(
0)}}\exp
[ik_z^{(0)}(z+z^{\prime })]-\exp (ik_z^{(0)}|z-z^{\prime }|)\right] ,
\label{eq:g13}
\\ [+3mm]
g_{31} & = & \frac{iK}{2\epsilon _{0}k_0^{2}}\left[ \frac{\epsilon
_{0}k_z^{(1)}-\epsilon _{1}k_z^{(0)}}{\epsilon _{0}k_z^{(1)}+\epsilon _{1}k_z^{(0)}}\exp
[ik_z^{(0)}(z+z^{\prime })]+\exp (ik_z^{(0)}|z-z^{\prime }|)\right] .
\label{eq:g31}
\end{eqnarray}

\noindent
We have used $k_z^{(j)} = iq_j = (K^2- \epsilon_j k_0^2) ^{1/2}$, $j=0$ ,$1$,
and $k_0=\omega/c$.
To determine the force acting on the particle, we also need the magnetic
field. This is found by the relationships ${\bf B}({\bf r})=-i/k\nabla
\times {\bf E}({\bf r})$. Then the time--averaged force obtained from
Maxwell's stress tensor $\overset{\leftrightarrow }
{{\bf T}}$~\cite{stratton41,jackson75} is

\begin{equation}
\langle {\bf F}\rangle =\int_{S}d^{2}r~\left\langle \overset{\leftrightarrow }{
{\bf T}}({\bf r})\right\rangle \cdot {\bf n}.  \label{eq:totalforce}
\end{equation}

\noindent
Equation~(\ref{eq:totalforce}) represents the flow of the time--average of
Maxwell's stress tensor
$\langle \overset{\leftrightarrow }{{\bf T}}\rangle$
across a surface $S$ enclosing the particle, ${\bf n}$ being the local
outward normal. The elements $T_{\alpha \beta }$ are~\cite{jackson75}

\begin{equation}
\left\langle T_{\alpha \beta }({\bf r})\right\rangle
=\frac{1}{8\pi }\left[ E_{\alpha
}E_{\beta }^{\ast }+B_{\alpha }B_{\beta }^{\ast }-\frac{1}{2}({\bf E}\cdot 
{\bf E}^{\ast }+{\bf B}\cdot {\bf B}^{\ast })\delta _{\alpha \beta }\right]
,(\alpha ,\beta =1,2,3).  \label{eq:maxwellstresstensor}
\end{equation}

\noindent
For dipolar particles, one can use instead of Maxwell's stress tensor the
expression given by Equation~(\ref{eq:chaumetfin}) directly. In fact, for
dielectric spheres of radii smaller than $5\times 10^{-2}\lambda $ there is
no appreciable difference between using Equation~(\ref{eq:chaumetfin}) or
Equation~(\ref{eq:maxwellstresstensor}) except at distances from the flat
substrate smaller than $10^{-3}\lambda $.

Figure~\ref{fig:dielec} shows the normalized $Z$--force for two glass
particles ($\epsilon =2.25$) at $\lambda =632.8$ $nm$, one with $a=10$ $nm$
(Figure~\ref{fig:dielec}(a)), and other with $a=100$ $nm$
(Figure~\ref{fig:dielec}(b)),
the flat interface is illuminated from the dielectric side
at $\theta_0 =42^{o}$, (the critical angle is $\theta_c =41.8^{o}$). Two
calculation procedures are shown: a multiple scattering evaluation of the
field via Equations~(\ref{eq:electricdip})--(\ref{eq:g31}) and then, either
use of Equation~(\ref{eq:chaumetfin}), integrated over all induced dipoles
(CDM--B), or Equation~(\ref{eq:totalforce}) (CDM--A). The normalization of the
forces has been carried out by dividing them by $\exp (-2qz)$. Thus, as seen
in these curves, the force tends, as $d$ increases, to the constant value
given by Equation~(\ref{eq:evanescent4}): $-(|T|^{2}/2)q\Re e \{ \alpha \}$.
The incident power is $1.19$ $mW$ distributed on a surface of $10$ $\mu
m^{2}$, then the force on a sphere of $a=10$ $nm$ is $2.7991\times 10^{-10}$ 
$pN$~\cite{chaumet00a}. We see, therefore, the effect on the vertical force
of the multiple interaction of the scattered wave with the substrate: as the
particle gets closer to the flat interface at which the evanescent wave is
created, the magnitude of the attractive force increases beyond the value
predicted by neglecting this interaction. As the distance to the surface
grows, the force tends to its value given by Equation~(\ref{eq:evanescent4})
in which no multiple scattering with the substrate takes place. Also, due to
the standing wave patterns that appear in the field intensity distribution
between the sphere and the substrate, the magnitude of this force oscillates
as $d$ varies. This is appreciated for larger particles
(Figure~\ref{fig:dielec}(b))
except for very small particles (Figure~\ref{fig:dielec}(a)),
whose scattering cross section is large enough to produce noticeable
interferences. On the other hand, the horizontal force on the particle is of
the form~given by Equation~(\ref{eq:evanescent3}) and always has the
characteristics of a scattering force.

\begin{figure}[t]
\begin{center}
\includegraphics*[draft=false,width=\linewidth]{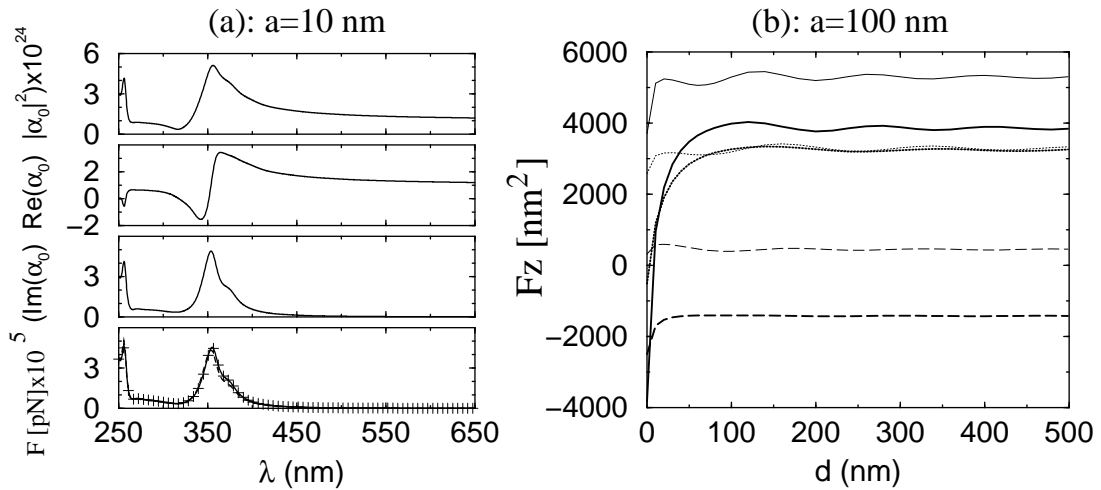}
\end{center}
\caption{ (a): From top to bottom: the first three curves represent the
polarizability of a silver sphere with radius $a=10$ $nm$ versus the
wavelength. The fourth curve is the force on this particle in free space.
Plain line: Mie calculation, dashed line: polarizability of
Eq.~(\ref{eq:alfa}),
symbol $+$: Dungey and Bohren's polarizability~\cite{dungey91}.
(b): Force along the $Z$
direction on a silver sphere with $a=100$ $nm$ versus distance $d$ with
$\protect\theta_0=50^o$ for the following wavelengths: Plain line: $\protect
\lambda=255$ $nm$, dashed line: $\protect\lambda=300$ $nm$, and dotted line: 
$\protect\lambda=340$ $nm$. Thin lines: $S$ polarization, thick lines: $P$
polarization. (From Ref.~\protect\cite{chaumet00b}). }
\label{fig:metallic}
\end{figure}

As regards a metallic particle, we notice that $\Re e \{ \alpha \}$ may
have negative values near plasmon resonances (Figure~\ref{fig:metallic}(a),
where we have plotted two models: that of Draine~\cite{draine88} and that of
Ref.~\cite{dungey91} (see also~\cite{chaumet00c})) and thus the gradient
force, or force along $OZ$, may now be repulsive, namely, positive
(Figure~\ref{fig:metallic}(b))~\cite{chaumet00c}.
We also observe that this force is
larger at the plasmon polariton resonance excitation ($\lambda =350$ $nm$).
We shall later return to this fact (Section~\ref{sec:resonances}). We next
illustrate how, no matter how small the particle is, the continuous approach
to the surface makes the multiple scattering noticeable.

\subsection{Corrugated Surfaces: Integral Equations for \newline
Light Scattering from Arbitrary Bodies}

\label{sec:corrugated}

At corrugated interfaces, the phenomenon of TIR is weakened, and the
contribution of propagating components to the transmitted field becomes
important, either from conversion of evanescent waves into radiating waves, or
due to the primary appearance of much propagating waves from scattering at
the surface defects. The size of the asperities is important in this
respect. However, TIR effects are still strong in slightly rough interfaces,
namely, those for which the defect size is much smaller than the wavelength.
Then the contribution of evanescent components is dominant on small
particles ({\it i.e}., of radius no larger than $0.1$ wavelengths). On the
other hand, the use of such particles as probes in near--field microscopy 
may allow high resolution of surface details as they scan above it. We shall
next study the resulting force signal effects due to corrugation, as a model
of photonic force microscopy under TIR conditions.

When the surface in front of the sphere is corrugated, finding the Green's
function components is not as straightforward as in the previous section. We
shall instead employ an integral method that we summarize next.

Let an electromagnetic field, with electric and magnetic vectors ${\bf E}
^{(inc)}({\bf r})$ and ${\bf H}^{(inc)}({\bf r})$, respectively, be incident on
a medium of permittivity $\epsilon $ occupying a volume $V$, constituted by
two scattering volumes $V_{1}$ and $V_{2}$, each being limited by a surface
$S_{1}$ and $S_{2}$, respectively. Let ${\bf r}^{<}$ be the position vector
of a generic point inside the volume $V_{j}$, and by ${\bf r}^{>}$ that of a
generic point in the volume $\hat{V}$, which is outside all volumes $V_{j}$.
The electric and magnetic vectors of a monochromatic field satisfy,
respectively, the wave equations, {\it i.e}., Equations~(\ref{eq:green1})
and~(\ref{eq:green2}).

The vector form of Green's theorem for two vectors ${\bf P}$ and ${\bf Q}$
well behaved in a volume $V$ surrounded by a surface $S$
reads~\cite{morsefeshbach53}

\begin{eqnarray}
\int_{V}d^{3}r~({\bf Q}\cdot \nabla \times \nabla \times {\bf P}-{\bf P}
\cdot \nabla \times \nabla \times {\bf Q}) & = & \nonumber \\
\int_{S}d^{2}r~({\bf P}\times
\nabla \times {\bf Q}-{\bf Q}\times \nabla \times {\bf P})\cdot {\bf n},
\label{eq:greentheorem}
\end{eqnarray}

\noindent
with ${\bf n}$ being the unit outward normal.

Let us now apply Equation~(\ref{eq:greentheorem}) to the vectors ${\bf P}=
\overset{\leftrightarrow }{{\cal G}}({\bf r},{\bf r}^{\prime })\cdot {\bf C}$
, (${\bf C}$ being a constant vector) and ${\bf Q}={\bf E}({\bf r})$. Taking
Equations~(\ref{eq:green1}) and~(\ref{eq:green5}) into account, we obtain

\begin{equation}
\int_{V}d^{3}r^{\prime }~{\bf E}({\bf r}^{\prime })\delta ({\bf r}-{\bf r}
^{\prime })=k^{2}\int_{V}d^{3}r^{\prime }~{\bf P}({\bf r}^{\prime })\cdot 
\overset{\leftrightarrow }{{\cal G}}({\bf r},{\bf r}^{\prime })-\frac{1}{
4\pi }{\bf S}_{e}({\bf r}),  \label{eq:ET1}
\end{equation}

\noindent
where ${\bf S}_{e}$ is

\begin{equation}
{\bf S}_{e}({\bf r})=\nabla \times \nabla \times \int_{S}d^{2}r^{\prime
}\left( {\bf E}({\bf r}^{\prime })\frac{\partial G({\bf r},{\bf r}^{\prime })
}{\partial {\bf n}}-G({\bf r},{\bf r}^{\prime })\frac{\partial {\bf E}({\bf r
}^{\prime })}{\partial {\bf n}}\right) .  \label{eq:ET2}
\end{equation}

\noindent
Equation~(\ref{eq:ET2}) adopts different forms depending on whether the
points ${\bf r}$ and ${\bf r}^{\prime }$ are considered in $V$ or in $\hat{V}
$. By means of straightforward calculations one obtains the following:

\begin{itemize}
\item If ${\bf r}$ and ${\bf r}^{\prime }$ belong to any of the volumes
$V_{j}$, $(j=1,2)$, namely, $V$ becomes either of the volumes $V_{j}$:

\begin{equation}
{\bf E}({\bf r}^{<})=k^{2}\int_{V_{j}}d^{3}r^{\prime }~{\bf P}({\bf r}
^{\prime })\cdot \overset{\leftrightarrow }{{\cal G}}({\bf r}^{<},{\bf r}
^{\prime })-\frac{1}{4\pi }{\bf S}_{j}^{(in)}({\bf r}^{<}),  \label{eq:ET3}
\end{equation}

where

\begin{eqnarray}
{\bf S}_{j}^{(in)}({\bf r}^{<}) & = & \nabla \times \nabla \times
\nonumber \\ [+3mm]
& &
\int_{S_{j}}d^{2}r^{\prime }\left( {\bf E}_{in}({\bf r}^{\prime })\frac{
\partial G({\bf r}^{<},{\bf r}^{\prime })}{\partial {\bf n}}-G({\bf r}^{<},
{\bf r}^{\prime })\frac{\partial {\bf E}_{in}({\bf r}^{\prime })}{\partial 
{\bf n}}\right) \, . \, \, \, \, \, \, \, \, \, \,
\label{eq:ET4}
\end{eqnarray}

In Equation~(\ref{eq:ET4}) ${\bf E}_{in}$ represents the limiting value of
the electric vector on the surface $S_{j}$ taken from inside the volume
$V_{j}$. Equation~(\ref{eq:ET3}) shows that the field inside each of the
scattering volumes $V_{j}$ does not depend on the sources generated in the
other volumes.

\item If ${\bf r}$ belongs to any of the volumes $V_{j}$, namely, $V$
becomes $V_{j}$, and ${\bf r}^{\prime }$ belongs to $\hat{V}$:

\begin{equation}
0={\bf S}_{ext}({\bf r}^{<}).  \label{eq:ET5}
\end{equation}

In Equation~(\ref{eq:ET5}) ${\bf S}_{ext}$ is

\begin{equation}
{\bf S}_{ext}({\bf r}^{<})=\sum_{j}{\bf S}_{j}^{(out)}({\bf r}^{<})-{\bf S}
_{\infty }({\bf r}^{<}),  \label{eq:ET6}
\end{equation}

where

\begin{eqnarray}
{\bf S}_{j}^{(out)}({\bf r}^{<}) & = & \nabla \times \nabla \times
\nonumber \\ [+3mm]
& &
\int_{S_{j}}d^{2}r^{\prime }\left( {\bf E}({\bf r}^{\prime })\frac{\partial
G({\bf r}^{<},{\bf r}^{\prime })}{\partial {\bf n}}-G({\bf r}^{<},{\bf r}
^{\prime })\frac{\partial {\bf E}({\bf r}^{\prime })}{\partial {\bf n}}
\right) \, . \, \, \, \, \, \, 
\label{eq:ET7}
\end{eqnarray}

In Equation~(\ref{eq:ET7}) the surface values of the electric vector are
taken from the volume $\hat{V}$. The normal ${\bf n}$ now points towards the
interior of each of the volumes $V_{j}$.

Also, ${\bf S}_{\infty }$ has the same meaning as Equation~(\ref{eq:ET7}),
the surface of integration now being a large sphere whose radius will
eventually tend to infinity. It is not difficult to see that $-{\bf S}
_{\infty }$ in Equation~(\ref{eq:ET6}) is $4\pi $ times the incident field
${\bf E}^{(inc)}({\bf r}^{<})$ ({\it cf}. Refs.~\cite{nieto-vesperinas91}
and~\cite{pattanayak76a,pattanayak76b}). Therefore Equation~(\ref{eq:ET5})
finally becomes

\begin{equation}
0={\bf E}^{(inc)}({\bf r}^{<})+\frac{1}{4\pi }\sum_{j}{\bf S}_{j}^{(out)}({\bf
r}^{<}).  \label{eq:ET8}
\end{equation}

Note that when Equation~(\ref{eq:ET8}) is used as a non--local boundary
condition, the unknown {\it sources} to be determined, given by the limiting
values of ${\bf E}({\bf r}^{\prime })$ and $\partial {\bf E}({\bf r}^{\prime
})/\partial {\bf n}$ on each of the surfaces $S_{j}$, ({\it cf}.
Equation~(\ref{eq:ET7})),
appear coupled to those corresponding sources on the other
surface $S_{k}$, $k\neq j$.

Following similar arguments, one obtains:

\item For ${\bf r}$ belonging to $\hat{V}$ and ${\bf r}^{\prime }$ belonging
to either volume $V_{j}$, $(j=1,2)$ namely, $V$ becoming $V_{j}$

\begin{equation}
0=k^2 \int_{V_{j}}d^{3}r^{\prime }~{\bf P}({\bf r}^{\prime
})\cdot \overset{\leftrightarrow }{{\cal G}}({\bf r}^{>},{\bf r}^{\prime })-
\frac{1}{4\pi }{\bf S}_{j}^{(in)}({\bf r}^{>}),  \label{eq:ET9}
\end{equation}

with ${\bf S}_{j}^{(in)}$ given by Equation~(\ref{eq:ET4}), this time
evaluated at ${\bf r}^{>}$.

\item For both ${\bf r}$ and ${\bf r}^{\prime }$ belonging to $\hat{V}$

\begin{equation}
{\bf E}({\bf r}^{>})={\bf E}^{(inc)}({\bf r}^{>})+\frac{1}{4\pi }\sum_{j}{\bf S
}_{j}^{(out)}({\bf r}^{>}),  \label{eq:ET10}
\end{equation}

Hence, the exterior field is the sum of the fields emitted from each
scattering surface $S_{j}$ $(j=1,2)$ with sources resulting from the
coupling involved in Equation~(\ref{eq:ET9}).
\end{itemize}

One important case corresponds to a penetrable, optically homogeneous,
isotropic, non--magnetic and spatially nondispersive medium (this applies for
a real metal or a pure dielectric). In this case, Equations~(\ref{eq:ET3})
and~(\ref{eq:ET9}) become, respectively,

\begin{eqnarray}
{\bf E}({\bf r}^{<}) &=&-\frac{1}{4\pi k_{0}^{2}\epsilon }\nabla \times
\nabla \times  \nonumber \\
&&\int_{S_{j}}d^{2}r^{\prime }\left( {\bf E}_{in}({\bf r}^{\prime })\frac{
\partial G^{(in)}({\bf r}^{<},{\bf r}^{\prime })}{\partial {\bf n}}-G^{(in)}(
{\bf r}^{<},{\bf r}^{\prime })\frac{\partial {\bf E}_{in}({\bf r}^{\prime })
}{\partial {\bf n}}\right) ,  \label{eq:ET11}
\end{eqnarray}

\begin{eqnarray}
0 &=&{\bf E}^{(inc)}({\bf r}^{<})+\frac{1}{4\pi k_{0}^{2}}\nabla \times \nabla
\times  \nonumber  \label{eq:ET12} \\
&&\sum_{j}\int_{S_{j}}d^{2}r^{\prime }\left( {\bf E}({\bf r}^{\prime })\frac{
\partial G({\bf r}^{<},{\bf r}^{\prime })}{\partial {\bf n}}-G({\bf r}^{<},
{\bf r}^{\prime })\frac{\partial {\bf E}({\bf r}^{\prime })}{\partial {\bf n}
}\right) ,
\end{eqnarray}

\noindent
whereas Equations~(\ref{eq:ET5}) and~(\ref{eq:ET10}) yield

\begin{eqnarray}
0 &=&\frac{1}{4\pi k_{0}^{2}}\nabla \times \nabla \times  \nonumber
\label{eq:ET13} \\
&&\int_{S_{j}}d^{2}r^{\prime }\left( {\bf E}_{in}({\bf r}^{\prime })\frac{
\partial G^{(in)}({\bf r}^{>},{\bf r}^{\prime })}{\partial {\bf n}}-G^{(in)}(
{\bf r}^{>},{\bf r}^{\prime })\frac{\partial {\bf E}_{in}({\bf r}^{\prime })
}{\partial {\bf n}}\right) ,
\end{eqnarray}

\begin{eqnarray}
{\bf E}({\bf r}^{>}) &=&{\bf E}^{(inc)}({\bf r}^{>})+\frac{1}{4\pi
k_{0}^{2}\epsilon }\nabla \times \nabla \times  \nonumber  \label{eq:ET14} \\
&&\sum_{j}\int_{S_{j}}d^{2}r^{\prime }\left( {\bf E}({\bf r}^{\prime })\frac{
\partial G({\bf r}^{>},{\bf r}^{\prime })}{\partial {\bf n}}-G({\bf r}^{>},
{\bf r}^{\prime })\frac{\partial {\bf E}({\bf r}^{\prime })}{\partial {\bf n}
}\right) .
\end{eqnarray}

\noindent
In Equations~(\ref{eq:ET11}) and~(\ref{eq:ET13}) $``in"$ means that the
limiting values on the surface are taken from inside the volume $V_{j}$;
note that this implies for both $G^{(in)}$ and ${\bf E}_{in}$ that $k=k_{0}
\sqrt{\epsilon }$.

The continuity conditions

\begin{equation}
{\bf n}\times \lbrack {\bf E}_{in}({\bf r}^{<})-{\bf E}({\bf r}
^{>})]=0,\,\,\,\,\,\,{\bf n}\times \lbrack {\bf H}_{in}({\bf r}^{<})-{\bf H}(
{\bf r}^{>})]=0 \, ,  \label{eq:ET15}
\end{equation}

\noindent
and the use of Maxwell's equations lead to (cf. Ref.~\cite{jackson75},
Section I.5, or Ref.~\cite{bornwolf99}, Section 1.1):

\begin{eqnarray}
\left. E _{in} ({\bf r}) \right| _{{\bf r} \in S_j^{(-)}} & = &
     \left. E ({\bf r}) \right| _{{\bf r} \in S_j^{(+)}} \, ,
\label{eq:continuity1}
\\ [+5mm]
     \left. \frac {\partial E _{in}({\bf r})} {\partial {\bf n}}
     \right|
     _{{\bf r} \in S_j^{(-)}} & = &
     \left. \frac {\partial E ({\bf r})} {\partial {\bf n}}
     \right|
     _{{\bf r} \in S_j^{(+)}} \, ,
\label{eq:continuity2}
\end{eqnarray}

\noindent
where $S_j^{(+)}$ and $S_j^{(-)}$ denote the surface profile when approached
from outside or inside the volume $V_j$, respectively.
Equations~(\ref{eq:continuity1}) and~(\ref{eq:continuity2}) permit to find both
${\bf E}$ and $\partial {\bf E}/\partial {\bf n}$
from either the pair Equations~(\ref{eq:ET13})
and~(\ref{eq:ET14}),
or, equivalently, from the pair Equations~(\ref{eq:ET11})
and~(\ref{eq:ET12}),
as both ${\bf r}^{>}$ and ${\bf r}^{<}$ tend to a point in
$S_{j}$. Then the scattered field outside the medium is given by the second
term of Equation~(\ref{eq:ET14}).

In the next section, we apply this theory to finding the near--field
distribution of light scattered from a small particle in front of a
corrugated dielectric surface when illumination is done from the dielectric
half--space at angles of incidence larger than the critical angle. The
non--local boundary conditions that we shall use are Equations~(\ref{eq:ET13})
and~(\ref{eq:ET14}).

\section{Photonic Force Microscopy of Surfaces with Defects}

\label{sec:PFM}

The {\em Photonic Force Microscope} (PFM) is a technique
in which one uses a probe particle trapped by a tweezer trap to image
soft surfaces. The PFM~\cite{ghislain93,florin96,wada00} was
conceived as a scanning probe device to measure ultrasmall forces, in the
range from a few to several hundredths $pN/nm$ with laser powers of some $mW$,
between colloidal particles~\cite{crocker94}, or in soft matter components
such as cell membranes~\cite{stout97} and protein or other macromolecule
bonds~\cite{smith96}. In such a system, a dielectric particle of a few
hundred nanometers, held in an optical
tweezer~\cite{ashkin86,clapp99,sugiura93,dogariu00},
scans the object surface. 
The spring constant
of the laser trap is three or four orders of magnitude smaller than that of
AFM cantilevers, and the probe position can be measured with a resolution of
a few nanometers within a range of some microseconds~\cite{florin96}. As in
AFM, surface topography imaging can be realized with a PFM by transducing
the optical force induced by the near field on the
probe~\cite{horber01bookproc}.
As in near--field scanning optical microscopy (NSOM\footnote{
NSOM is also called SNOM, abbreviation for scanning near--field optical
microscopy.}) \cite{pohl93}, the
resolution is given by the size of the particle and its proximity to the
surface. It is well known, however~\cite{nieto-vesperinas91,greffet97} that
multiple scattering effects and artifacts, often hinder NSOM images so that
they do not bear resemblance to the actual topography . This has constituted
one of the leading basic problems in NSOM~\cite{hecht96}. Numerical
simulations~\cite{yo01b,yo01c,yo99b,yo00,yo01bookproc} based on the theory
of Section~\ref{sec:corrugated} show that detection of the optical force on
the particle yields topographic images, and thus they provide a method of
prediction and interpretation for monitoring the force signal variation with
the topography, particle position and illumination conditions. This
underlines the fundamentals of the PFM operation. An important feature is
the signal enhancement effects arising from the excitation of {\it Mie
resonances} of the particle, which we shall discuss next. This allows to
decrease its size down to the nanometric scale, thus increasing resolution
both of force magnitudes and spatial details.

\subsection{Nanoparticle Resonances}

\label{sec:resonances}

Electromagnetic eigenmodes of small particles are of importance in several
areas of research. On the one hand, experiments on the linewidth of surface
plasmons in metallic particles~\cite{klar98} and on the evolution of their
near fields, both in isolated particles and in arrays~\cite{krenn99}, seek a
basic understanding and possible applications of their optical properties.

Mie resonances of particles are often called {\it morphology--dependent
resonances} ({\it MDR}). They depend on the particle shape, permittivity,
and the {\it size parameter}: $x=2\pi a/\lambda $. In dielectric particles,
they are known as {\it whispering--gallery modes} ({\it WGM})~\cite
{owen81,barber82,benincasa87,hill88,barber88,barber90}. On the other hand, in
metallic particles, they become {\it surface plasmons} ({\it SPR}), coming
from electron plasma oscillations~\cite{raether88}. All these resonances are
associated to surface waves which exponentially decay away from the particle
boundary.

Morphology--dependent resonances in dielectric particles are interpreted as
waves propagating around the object, confined by total internal reflection,
returning in phase to the starting point. A {\it Quality factor} is also
defined as $Q=2\pi $ (Stored energy) $/$ (Energy lost per cycle) $=\omega
_{0}/\delta \omega $, where $\omega _{0}$ is the resonace frequency and
$\delta \omega $ the resonance full width. The first theoretical studies of 
{\it MDR} were performed by Gustav Mie, in his well--known scattering theory
for spheres. The scattered field, both outside and in the particle, is
decomposed into a sum of partial waves. Each partial wave is weighted by a
coefficient whose poles explain the existence of peaks at the scattering
cross section. These poles correspond to complex frequencies, but true
resonances ({\it i.e}.,the real values of frequency which produce a finite
for the coefficient peaks) have a size parameter value close to the real
part of the complex poles. The imaginary part of the complex frequency
accounts for the width of the resonance peak. {\it MDR}'s are classified by
three integer numbers: one related to the partial wave ({\it order number}),
another one which accounts for the several poles that can be present in the
same coefficient ({\it mode number}), and a third one accounting for the
degeneration of a resonance ({\it azimuthal mode number}). In the first
experimental check at optical frequencies, the variation of the radiation
pressure (due to {\it MDR}) on highly transparent, low--vapor--pressure
silicone oil drops (index $1.4-1.53$) was measured by Ashkin~\cite{ashkin77}.
The drops were levitated by optical techniques and the incident beam was
focused at either the edge or the axis of the particles showing the creeping
nature of the surface waves.

It is important to note,
as regards resonances, the enhanced directional scattering
effects such as the {\it Glory}~\cite{bryant66,fahlen68,khare77} found in
water droplets.
The Glory theory accounts for the backscattering intensity
enhancements found in water droplets. These enhancements are associated with
rays grazing the surface of the droplet, involving hundreds of
circumvolutions (surface effects). Axial rays (geometrical effects) also
contribute. They have been observed in large particle sizes ($x>10^{2}$) and
no Glory effects have been found for sizes in the range $x\sim 1$. These
backscattering intensity enhancements cannot be associated to a unique
partial wave, but to a superposition of several partial waves.

\subsection{Distributions of Forces on Dielectric Particles \newline
Over Corrugated Surfaces Illuminated Under TIR}

\label{sec:forcedielec}

\begin{figure}[t]
\begin{center}
\includegraphics*[draft=false,width=\linewidth]{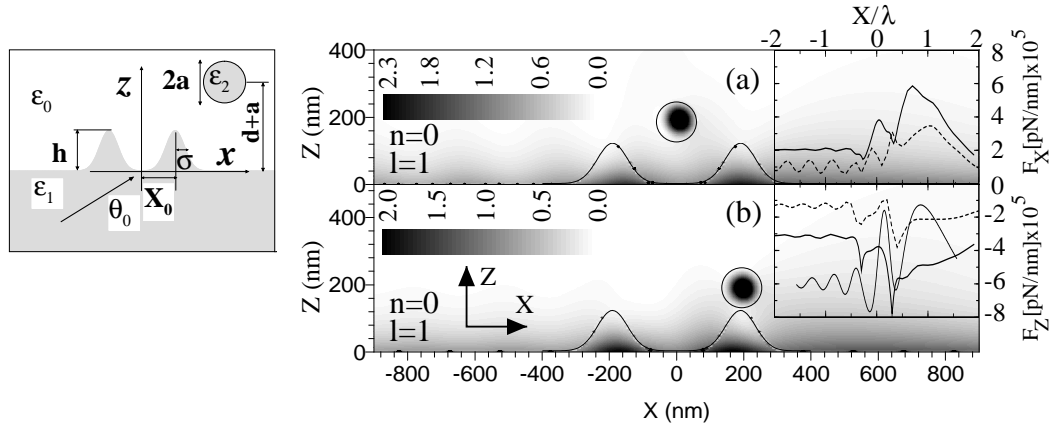}
\end{center}
\caption{ Left figure: Scattering geometry. Insets: Force curves on a
silicon cylinder with $a=60$ $nm$ scanned at $d=132.6$ $nm$. (a) Horizontal
force. (b) Vertical force. Solid line: $\protect\lambda=638$ $nm$ (on
resonance). Broken line: $\protect\lambda=538$ $nm$ (off resonance). Thin
solid line in (b): $|H/H_ o|^2$ at $z=d+a$ in absence of particle. Peak
value: $|H/H_ o|^2=0.07$. Bottom figures: Spatial distribution $|H/H_ o|^2$
in this configuration. The cylinder center is placed at: $(0, 192.6)$ $nm$
(Fig.~\ref{fig:ol1}(a)), and at: $(191.4, 192.6)$ $nm$ (Fig.~\ref{fig:ol1}(b)).
The wavelength ($\protect\lambda=638$ $nm$) excites the $(n,l)$ Mie
resonance. (From Ref.~\protect\cite{yo01b}). }
\label{fig:ol1}
\end{figure}

We now model a rough interface separating a dielectric of permittivity
$\epsilon _{1}=2.3104$, similar to that of glass, from air. We have addressed
(Figure~\ref{fig:ol1}, left) the profile consisting of two protrusions described
by $z=h[\exp (-(x-X_{0})^{2}/\sigma ^{2})+\exp (-(x+X_{0})^{2}/\sigma ^{2})]$
on a plane surface $z=0$. (It should be noted that in actual experiments,
the particle is immersed in water, which changes the particle's
relative refractive index weakly. But the phenomena shown here will remain, 
with the interesting features now occurring at slightly different
wavelengths.) Illumination, linearly polarized, is done from the dielectric
side under TIR (critical angle $\theta _{c}=41.14^{o}$) at $\theta
_{0}=60^{o}$ with a Gaussian beam of half--width at half--maximum $W=4000$ $nm$
at wavelength $\lambda $ (in air).
For the sake of computing time and memory, the calculation is done in two
dimensions (2D). This retains the main physical features of the full
3D configuration, as far as multiple interaction of the field with the
surface and the probe is concerned~\cite{lester99}. The particle is then a
cylinder of radius $a$, permittivity $\epsilon _{2}$, and axis $OY$, whose
center moves at constant height $z=d+a$. Maxwell's stress tensor is used to
calculate the force on the particle resulting from the scattered near--field
distribution created by multiple interaction of light between the surface
and the particle. Since the configuration is 2D, the incident power and the
force are expressed in $mW/nm$ and in $pN/nm$, respectively, namely, as
power and force magnitudes per unit length (in $nm$) in the transversal
direction, {\it i.e}., that of the cylinder axis. We shall further discuss
how these magnitudes are consistent with three--dimensional (3D) experiments.

A silicon cylinder of radius $a=60$ $nm$ in front of a flat dielectric
surface with the same value of $\epsilon _{1}$ as considered here, has a Mie
resonance excited by the transmitted evanescent wave at $\lambda =638$ $nm$
($\epsilon _{2}=14.99+i0.14$)~\cite{yo00}. Those eigenmodes are characterized
by $n=0$, $l=1$, for $p$ polarization, and $n=1$, $l=1$, for $s$
polarization. We consider the two protrusion interface (Figure~\ref{fig:ol1},
left). Insets of
Figures~\ref{fig:ol1}(a) and~\ref{fig:ol1}(b), corresponding to a $p$--polarized
incident beam, show the electromagnetic force on the particle as it scans
horizontally above the flat surface with two protrusions with parameters
$\sigma =63.8$ $nm$, $h=127.6$ $nm$ and $X_{0}=191.4$ $nm$, both at resonant
$\lambda $ and out of resonance ($\lambda =538$ $nm$). The particle scans at
$d=132.6$ $nm$. Inset (a) shows the force along the $OX$ axis. As seen, the
force is positive and, at resonance, it has two remarkable maxima
corresponding to the two protrusions, even though they appear slightly
shifted due to surface propagation of the evanescent waves transmitted under
TIR, which produce the Goos--H\"{a}nchen shift of the reflected beam. The
vertical force on the particle, on the other hand, is negative, namely
attractive, ({\it cf}. Inset (b)), and it has two narrow peaks at $x$ just
at the position of the protrusions. The signal being again remarkably
stronger at resonant illumination. Similar force signal enhancements are
observed for $s$--polarization. In this connection, it was recently found
that this attractive force on such small dielectric particles monotonically
increases as they approach a dielectric flat inteface~\cite{chaumet00a}.

\begin{figure}[t]
\begin{center}
\includegraphics*[draft=false,width=\linewidth]{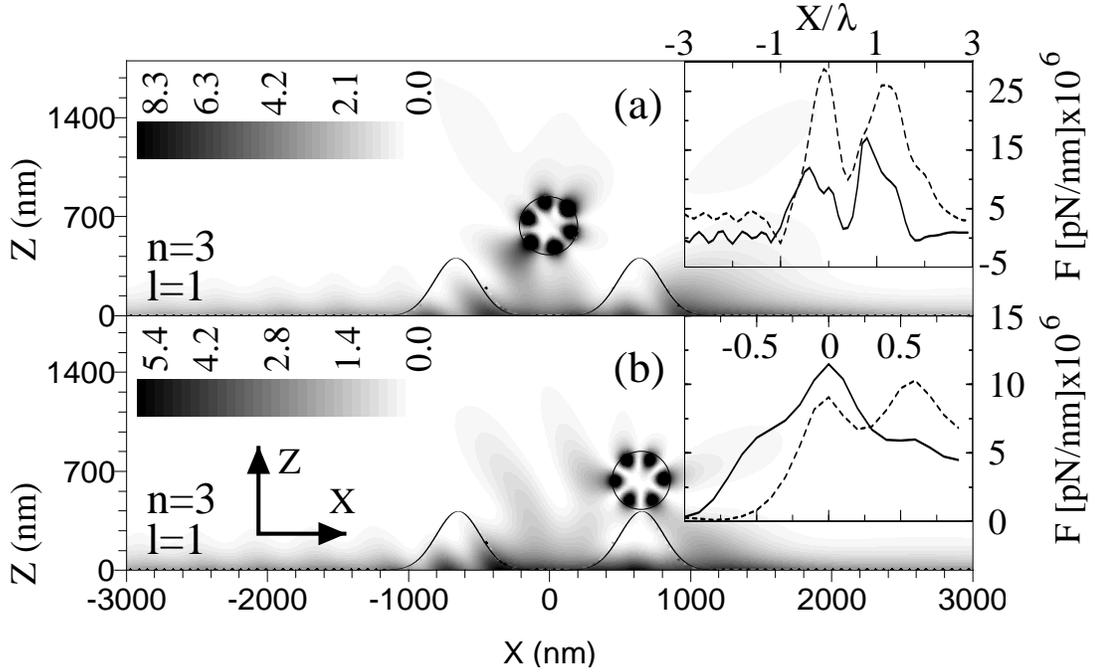}
\end{center}
\caption{ Insets: Force on a silicon cylinder with $a=200$ $nm$ scanned at
$d=442$ $nm$. (a) $\protect\lambda=919$ $nm$ (on resonance). (b): $\protect
\lambda=759$ $nm$ (off resonance). Solid line: Vertical force. Broken line:
Horizontal force. Figures: $|E/E_ o|^2$ in this configuration. The cylinder
center is placed at: $(0, 642)$ $nm$ (Fig.~\ref{fig:ol2}(a)), and at: $(638,
642)$ $nm$ (Fig.~\ref{fig:ol2}(b)). The wavelength ($\protect\lambda=919$
$nm$) excites the $(n,l)$ Mie resonance. (From Ref.~\protect\cite{yo01b}). }
\label{fig:ol2}
\end{figure}

It should be remarked that, by contrast and as
expected~\cite{nieto-vesperinas91,greffet97},
the near field--intensity distribution for
the magnetic field $H$, normalized to the incident one $H_{0}$, has many
more components and interference fringes than the force signal, and thus, 
the resemblance of its image with the interface topography is worse.
This is shown in Inset (b) (thin solid line), where we have plotted this
distribution in the absence of particle at $z=d+a$ for the same illumination
conditions and parameters as before. This is the one that ideally the
particle scan should detect in NSOM.

It is also interesting to investigate the near--field intensity distribution
map. Figures~\ref{fig:ol1}(a) and~\ref{fig:ol1}(b) show this for the magnetic
field $H$ in $p$ polarization for resonant illumination, $\lambda =638$ $nm$,
at two different positions of the cylinder, which correspond to $x=0$ and
$191.4$ $nm$, respectively. We notice, first, the strong field concentration
inside the particle, corresponding to the excitation of the $(n=0,l=1)$
--eigenmode. When the particle is over one bump, the variation of the near
field intensity is larger in the region closer to it, this
being responsible for the stronger force signal on the particle
at this position.
Similar results are observed for $s$--polarized
waves, in this case, the $(n=1,l=1)$--eigenmode of the cylinder is excited
and one can appreciate remarkable fringes along the whole interface due to
interference of the surface wave, transmitted under TIR, with scattered
waves from both the protrusions and the cylinder.

An increase of the particle size yields stronger force signals at the
expense of losing some resolution. Figures~\ref{fig:ol2}(a)
and~\ref{fig:ol2}(b)
show the near electric field intensity distribution for $s$ polarization at
two different positions of a cylinder with radius $a=200$ $nm$, the
parameters of the topography now being $\sigma =212.7$ $nm$, $h=425.3$ $nm$
and $X_{0}=\pm 638$ $nm$. The distance is $d=442$ $nm$. The resonant
wavelength is now $\lambda =919$ $nm$ ($\epsilon _{2}=13.90+i0.07$). Insets
of Figures~\ref{fig:ol2}(a) and ~\ref{fig:ol2}(b) illustrate the force
distribution as the cylinder moves along $OX$. The force peaks, when
the resonant wavelength is considered, are positive, because now,
the scattering force
on this particle of larger scattering cross section is greater than the
gradient force. They also appear shifted with respect to the protrusion
positions, once again due to surface travelling waves under TIR. There are
weaker peaks, or absence of them at non--resonant $\lambda $. Similar results
occur for a $p$--polarized beam at the resonant wavelength $\lambda =759$ $nm$
($\epsilon _{2}=13.47+i0.04$). For both polarizations the $(n=3,l=1)$ Mie
eigenmode of the cylinder is now excited. The field distribution is
well--localized inside the particle and it has the characteristic standing
wave structure resulting from interference between counterpropagating
whispering--gallery modes circumnavegating the cylinder surface. It is
remarkable that this structure appears as produced by the excitation of
propagating waves incident on the particle~\cite{yo00}, these being due to
the coupling of the incident and the TIR surface waves with radiating
components of the transmitted field, which are created from scattering with
the interface protrusions. Although not shown here, we should remark that
illumination at non--resonant wavelengths do not produce such a field
concentration within the particle, then the field is extended throughout the
space, with maxima attached to the flat portions of the interface
(evanescent wave) and along certain directions departing from the
protrusions (radiating waves from scattering at these surface defects).

Evanescent components of the electromagnetic field and multiple scattering
among several objects are often difficult to handle in an experiment.
However, there are many physical situations that involve these phenomena.
In this section we have seen that the use of the field inhomogeneity, combined
with (and produced by) morphology--dependent resonances and multiple scattering,
permit to imaging a surface with defects. Whispering--gallery modes in
dielectric
particles,
on the other hand,
produce also evanescent fields on the particle surface
which enhance the strength of the force signal.
The next section is aimed to study metallic particles under the same situation
as previously discussed,
exciting now plasmon resonances on the objects.

\subsection{Distributions of Forces on Metallic Particles \newline
Over Corrugated Surfaces Illuminated Under TIR}

\label{sec:forcemetal}

Dielectric particles suffer intensity gradient forces under light
illumination due to radiation pressure, which permit one to hold and
manipulate them by means of optical tweezers~\cite{ashkin77} in a
variety of applications such as
spectroscopy \cite{sasaki91,misawa92,misawa91},
phase transitions in polymers 
\cite{hotta98},
and light force microscopy of cells \cite{pralle99,pralle98} and
biomolecules \cite{smith96}. Metallic particles, however, were initially
reported to suffer {\em repulsive} electromagnetic scattering forces due to
their higher cross sections~\cite{ashkin92}, although later~\cite{svoboda94}
it was shown that nanometric metallic particles (with diameters smaller than 
$50$ $nm$) can be held in the focal region of a laser beam.
Further, it was demonstrated in an experiment~\cite{sasaki00}
that metallic particles illuminated by an evanescent wave created under TIR
at a substrate, experience a vertical attractive force towards the plate,
while they are pushed horizontally in the direction of propagation of the
evanescent wave along the surface. Forces in the $fN$ range were measured.

\begin{figure}[t]
\begin{center}
\includegraphics*[draft=false,width=9.5cm]{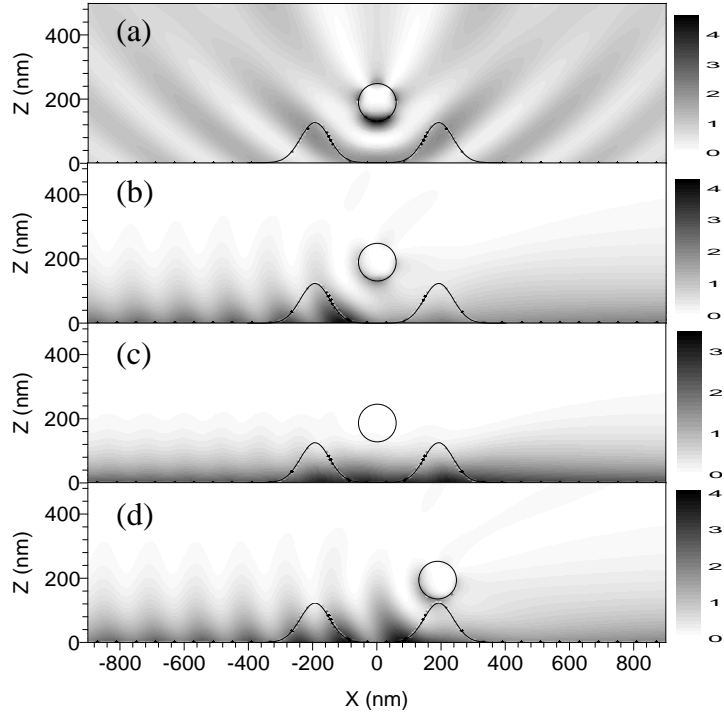}
\end{center}
\caption{$|H/H_ o|^2$, $P$ polarization, from a silver cylinder with $a=60$
$nm$ immersed in water, on a glass surface with defect parameters
$X_0=\pm191.4$ $nm$, $h=127.6$ $nm$
and $\protect\sigma=63.8$ $nm$, at distance
$d=132.6$ $nm$. Gaussian beam incidence with $W=4000$
$nm$.~\ref{fig:prb3}(a):
$\protect\lambda =387$ $nm$ (on resonance),
$\protect\theta_ o=0^o$.~\ref{fig:prb3}(b):
$\protect\lambda =387$ $nm$ (on resonance),
$\protect\theta_o=66^o$.~\ref{fig:prb3}(c):
$\protect\lambda =316$ $nm$ (off
resonance), $\protect\theta_o=66^o$.~\ref{fig:prb3}(d):
$\protect\lambda=387$ $nm$ (on resonance),
$\protect\theta_o=66^o$. The cylinder center is
placed at $(0, 192.6)$ nm in~\ref{fig:prb3}(a),~\ref{fig:prb3}(b)
and~\ref{fig:prb3}(c),
and at $(191.4, 192.6)$ nm in~\ref{fig:prb3}(d).
(From Ref.~\protect\cite{yo01c}). }
\label{fig:prb3}
\end{figure}

Plasmon resonances in metallic particles are not so efficiently excited as
mor-phology--dependent resonances in non--absorbing high--refractive--index
dielectric particles ({\it e.g}., see Refs.~\cite{yo01a,yo00}) under
incident evanescent waves. The distance from the particle to the surface
must be very small to avoid the evanescent wave decay normal to the
propagation direction along the surface. In this section we address the same
configuration as before using water as the immersion medium. The critical
angle for the glass--water interface is $\theta _{c}=61.28^{o}$. A silver
cylinder of radius $a$ at distance $d+a$ from the flat portion of the
surface is now studied.

In Figure~\ref{fig:prb3} we plot the near--field intensity distribution
$|H/H_{0}|^{2}$ corresponding to the configuration of inset in
Figure~\ref{fig:ol1}.
A silver cylinder of radius $a=60$ $nm$ scans at constant
distance $d=162.6$ $nm$ above the interface. The system is illuminated by a
$p$--polarized Gaussian beam ($W=4000$ $nm$) at $\theta _{0}=0^{o}$ and
$\lambda =387$ $nm$ ($\epsilon _{2}=-3.22+i0.70$). The surface protrusions
are positioned at $X_{0}=\pm 191.4$ $nm$ with height $h=127.6$ $nm$ and
$\sigma =63.8$ $nm$. Figure~\ref{fig:prb3}(a) shows the aforementioned
distribution when the particle is centered between the protrusions. The
plasmon resonance is excited as manifested by the field enhancement on the
cylinder surface, which is higher in its lower portion. At this resonant
wavelength, the main Mie coefficient contributor is $n=2$, which can also be
deduced from the interference pattern formed along the particle surface: the
number of lobes must be $2n$ along this surface~\cite{owen81}.
Figure~\ref{fig:prb3}(b)
shows the same situation but with $\theta _{0}=66^{o}$. The
field intensity close to the particle is higher in Figure~\ref{fig:prb3}(a)
because in Figure~\ref{fig:prb3}(b) the distance $d$ is large enough to
obliterate the resonance excitation due to the
decay of the evanescent wave created by TIR~\cite{yo01a}. However, the field
intensity is markedly different from the one shown in
Figure~\ref{fig:prb3}(c),
in which the wavelength has been changed to $\lambda =316$ $nm$
($\epsilon _{2}=0.78+i1.07$) so that there is no particle resonance excitation
at all. Figure~\ref{fig:prb3}(d) shows the same situation as in
Figure~\ref{fig:prb3}(b)
but at a different $X$--position of the particle.
In Figure~\ref{fig:prb3}(c),
the interference in the scattered near--field due to the
presence of the particle is rather weak; the field distribution is now seen
to be mainly concentrated at low $z$ as an evanescent wave travelling along
the interface, and this distribution does not substantially change as the
particle moves over the surface at constant $z$. By contrast, in
Figures~\ref{fig:prb3}(b)
and~\ref{fig:prb3}(d) the intensity map is strongly perturbed by
the presence of the particle. As we shall see, this is the main reason due
to which optical force microscopy is possible at resonant conditions with
such small metallic particles used as nanoprobes, and not so efficient at
non--resonant wavelengths. In connection with these intensity maps
({\it cf}. Figures~\ref{fig:prb3}(b) and~\ref{fig:prb3}(d)), we should point out
the interference pattern on the left side of the cylinder between the
evanescent wave and the strongly reflected waves from the particle, that in
resonant conditions behaves as a strongly radiating
antenna~\cite{yo01a,yo00,krenn99}.
This can also be envisaged as due to the much larger scattering cross
section of the particle on resonance, hence reflecting backwards higher
intensity and thus enhancing the interference with the evanescent incident
field. The fringe spacing is $\lambda /2$ ($\lambda $ being the
corresponding wavelength in water). This is explained as follows: The
interference pattern formed by the two evanescent waves travelling on the
surface opposite to each other, with the same amplitude and no dephasing, is
proportional to $\exp (-2\kappa z)\cos ^{2}(n_{1}k_{0}\sin \theta _{0}x)$,
with $\kappa =(n_{1}^{2}\sin ^{2}\theta _{0}-n_{0}^{2})^{1/2}$. The distance
between maxima is $\Delta x=\lambda /(2n_{1}\sin \theta _{0})$. For the
angles of incidence used in this work under TIR ($\theta _{0}=66^{0}$ and
$72^{0}$), $\sin \theta _{0}\approx 0.9$, and taking into account the
refractive indices of water and glass, one can express this distance as
$\Delta x\approx \lambda /2n_{0}$. The quantity $\Delta x$ is similar to the
fringe period below the particle in Figure~\ref{fig:prb3}(a), now attributted
to the interference between two opposite travelling plane waves, namely, the
one transmitted through the interface and the one reflected back from the
particle.

\begin{figure}[h]
\begin{center}
\includegraphics*[draft=false,width=\linewidth]{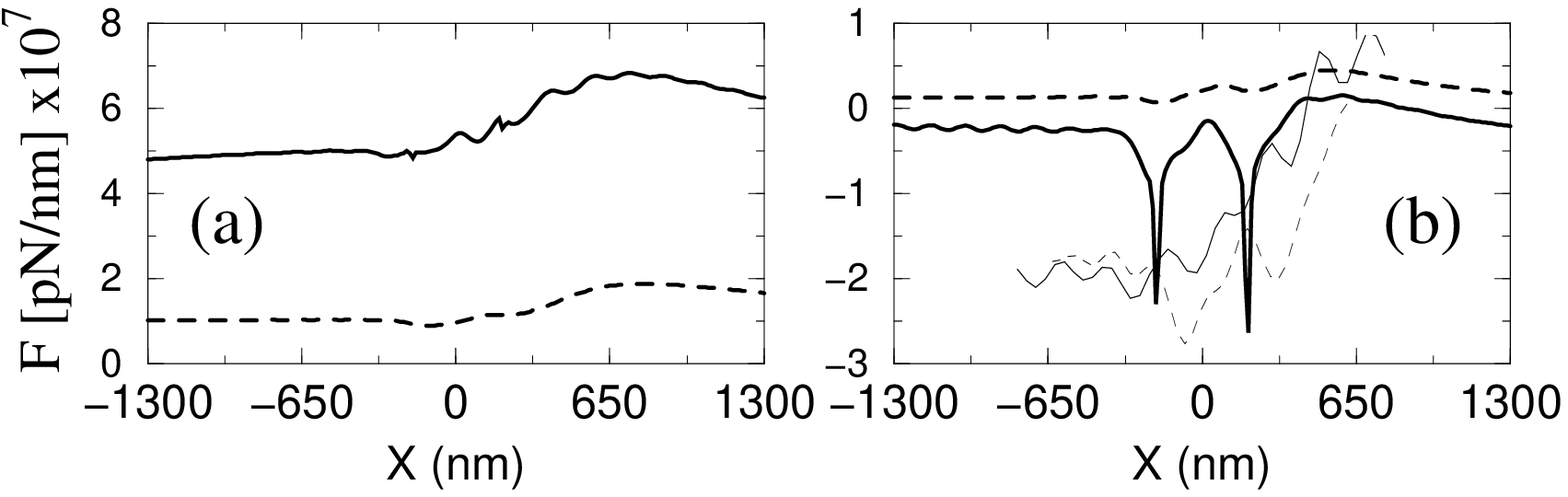}
\end{center}
\caption{ Force on a silver cylinder with $a=60$ $nm$ immersed in water,
scanned at constant distance $d=132.6$ $nm$ on a glass surface with defect
parameters $X_0=\pm 191.4$ $nm$, $h=127.6$ $nm$ and $\protect\sigma=63.8$
$nm$ along $OX$. The incident field is a $p$--polarized Gaussian beam with
$W=4000$ $nm$ and $\protect\theta_ 0=66^o$.~\ref{fig:prb5}(a): Horizontal
force.~\ref{fig:prb5}(b): Vertical force. Solid curves: $\protect\lambda=
387$ $nm$ (on resonance), broken curves: $\protect\lambda= 316$ $nm$ (off
resonance). Thin lines in~\ref{fig:prb5}(b) show $|H/H_0|^2$ (in arbitrary
units), averaged on the perimeter of the cylinder cross section, while it
scans the surface. The actual magnitude of the intensity in the resonant
case is almost seven times larger than in the non--resonant one.
(From Ref.~\protect\cite{yo01c}). }
\label{fig:prb5}
\end{figure}

\begin{figure}[t]
\begin{center}
\includegraphics*[draft=false,width=11cm]{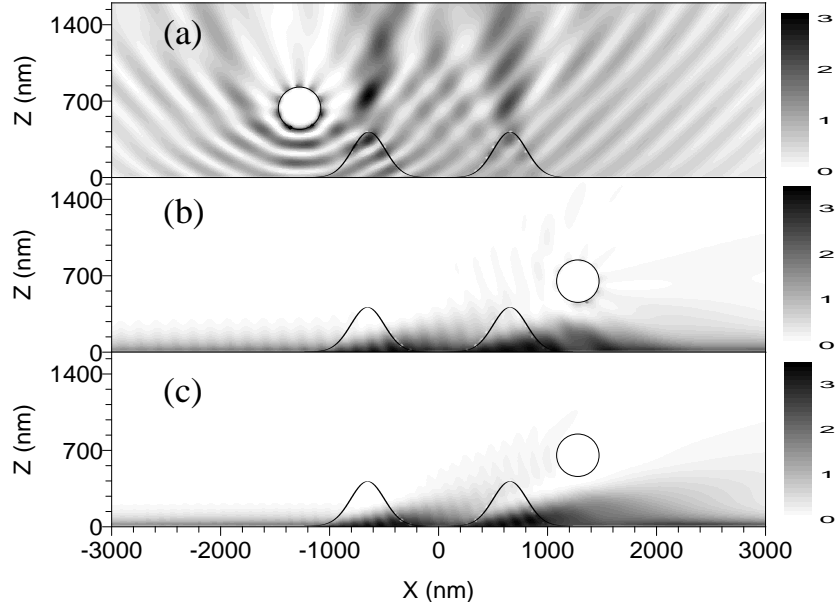}
\end{center}
\caption{$|H/H_ o|^2$ for $P$ polarization for a silver cylinder with $a=200$
$nm$ immersed in water, on a glass surface with parameters $X_0=\pm 638$ $nm$,
$h=425.3$ $nm$ and $\protect\sigma=212.7$ $nm$, at distance $d=442$ $nm$.
Gaussian beam incidence with $W=4000$ $nm$.~\ref{fig:prb8}(a): $\protect
\lambda =441$ $nm$ (on resonance), $\protect\theta_ o=0^o$ and the cylinder
center placed at $(-1276, 642)$ $nm$.~\ref{fig:prb8}(b): $\protect\lambda
=441$ $nm$ (on resonance), $\protect\theta_o=66^o$ and the cylinder center
placed at $(1276, 642)$ $nm$.~\ref{fig:prb8}(c): $\protect\lambda =316$ $nm$
(off resonance), $\protect\theta_o=66^o$ and the cylinder center placed at
$(1276, 642)$ $nm$. (From Ref.~\protect\cite{yo01c}). }
\label{fig:prb8}
\end{figure}

Figure~\ref{fig:prb5} shows the variation of the Cartesian components of the
electromagnetic force ($F_x$, Fig.~\ref{fig:prb5}(a) and $F_z$,
Fig.~\ref{fig:prb5}(d))
on scanning the particle at constant distance $d$ above the
interface, at either plasmon resonance excitation ($\lambda=387$ $nm$, solid
lines), or off resonance ($\lambda=316$ $nm$, broken lines). The incident
beam power (per unit length) on resonance is $3.9320 \times 10^{-6}$ $mW/nm$,
and $3.9327\times 10^{-6}$ $mW/nm$ at $\lambda=316$ $nm$. The incidence is
done with a $p$--polarized Gaussian beam of $W=4000$ $nm$ at $\theta_0=66^o$.
It is seen from these curves that the force distributions resembles the
surface topography on resonant conditions with a signal which is remarkably
larger than off--resonance. This feature is specially manifested in the $Z$
component of the force, in which the two protrusions are clearly
distinguished from the rest of interference ripples, as explained above.
Figure~\ref{fig:prb5}(b) also shows (thin lines) the scanning that
conventional near field microscopy would measure in this configuration,
namely, the normalized magnetic near field intensity, averaged on the
cylinder cross section. These intensity curves are shown in arbitrary units,
and in fact the curve corresponding to plasmon resonant conditions is almost
seven times larger than the one off--resonance. The force curves show, on the
one hand, that resonant conditions also enhance the contrast of the surface
topography image. Thus, the images obtained from the electromagnetic force
follows more faithfully the topography than that from the near field
intensity. This is a fact also observed with other profiles, including
surface--relief gratings. When parameter $h$ is inverted, namely, the
interface profile on the left in Fig.~\ref{fig:ol1}, then the vertical
component of the force distribution presents inverted the contrast. On the
whole, one observes from these results that both the positions and sign of
the defect height can be distinguished by the optical force scanning.

\begin{figure}[h]
\begin{center}
\includegraphics*[draft=false,width=\linewidth]{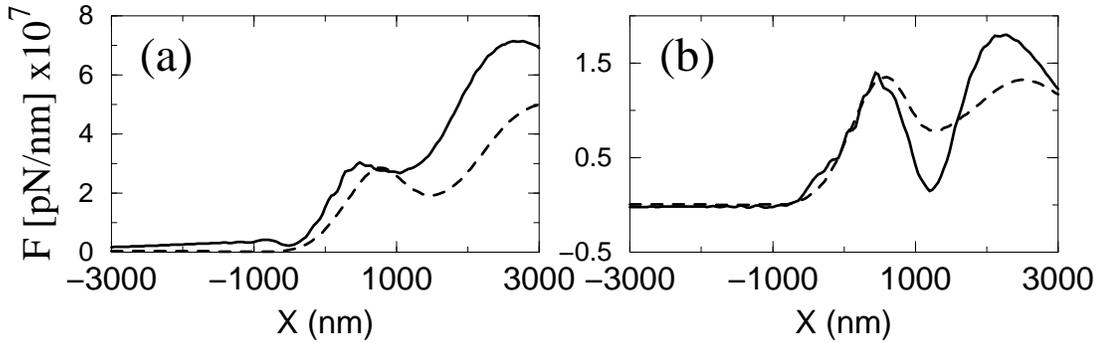}
\end{center}
\caption{ Force on a silver cylinder with $a=200$ $nm$ immersed in water,
scanned at constant distance on a glass surface with parameters $X_0=\pm 638$
$nm$, $h=425.3$ and $\protect\sigma=212.7$ $nm$ along $OX$. The incident
field is a $p$--polarized Gaussian beam with $W=4000$ $nm$ and
$\protect\theta_ 0=66^o$.~\ref{fig:prb9}(a):
Horizontal force.~\ref{fig:prb9}(b):
Vertical force. Solid curves: $\protect\lambda= 441$ $nm$ (on resonance),
broken curves: $\protect\lambda= 316$ $nm$ (off resonance). Thin solid
curves: $\protect\lambda= 441$ $nm$ (on resonance) at $\protect\theta_
0=72^o $, thin broken curves: $\protect\lambda= 316$ $nm$ (off resonance) at 
$\protect\theta_ 0=72^o$. (From Ref.~\protect\cite{yo01c}). }
\label{fig:prb9}
\end{figure}

Figure~\ref{fig:prb8} displays near--field intensity maps for a larger particle
($a=200$ $nm$).
Figure~\ref{fig:prb8}(a)
corresponds to $\theta _{0}=0^{o}$ and a resonant wavelength
$\lambda =441$ $nm$ ($\epsilon _{2}=-5.65+i0.75$), with the particle being
placed on the left of both protrusions. Figure~\ref{fig:prb8}(b) corresponds
to $\theta _{0}=66^{o}$ (TIR illumination conditions), at the same resonant
wavelength, the particle now being on the right of the protrusions.
Figure~\ref{fig:prb8}(c)
corresponds to $\theta _{0}=66^{o}$ (TIR incidence), at the
no resonant wavelength $\lambda =316$ $nm$ ($\epsilon _{2}=0.78+i1.07$), the
particle being placed at the right of the protrusions. The incident beam is
$p$--polarized with $W=4000$ $nm$. The surface protrusions are positioned at
$X_{0}=\pm 638$ $nm$ with height $h=425.3$ $nm$ and $\sigma =212.7$ $nm$. All
the relevant size parameters are now comparable to the wavelength, and hence
to the decay length of the evanescent wave. That is why now the plasmon
resonance cannot be highly excited. When no resonant wavelength is used, the
intensity interference fringes due to the presence of the particle are
weaker. On the other hand, Figure~\ref{fig:prb8}(a) shows the structure of the
near--field scattered under $\theta _{0}=0^{o}$. There are three objects that
scatter the field: the two protrusions and the particle. They create an
inteference pattern with period $\lambda /2$ (with $\lambda $ being the
wavelength in water). Besides, the particle shows an inteference pattern
around its surface due to the two counterpropagating plasmon waves which
circumnavegate it~\cite{yo01a,yo00}. The number of lobes along the surface
is nine, which reflects that the contribution to the field enhancement at
this resonant wavelength comes from Mie's coefficients $n=5$ and $n=4$.
Figure~\ref{fig:prb8}(b) shows weaker excitation of the same
plasmon resonance under TIR conditions. Now,
the interference pattern at the incident side of the configuration is also
evident. This
pattern again has a period $\lambda /2$ ($\lambda $ being the wavelength in
water). If non--resonant illumination conditions are used, the particle is
too far from the surface to substantially perturb the transmitted evanescent
field,
then the intensity distribution of this field remains closely
attached to the interface, and it is
scattered by the surface protrusions. The
field felt by the particle in this situation is not sufficient to yield a
well--resolved image of the surface topography, as shown next for this same
configuration.

Figure~\ref{fig:prb9} shows the components of the force ($F_{x}$,
Figure~\ref{fig:prb9}(a)
and $F_{z}$, Figure~\ref{fig:prb9}(b)) for either plasmon
excitation conditions ($\lambda =441$ $nm$, solid lines), or off--resonance
($\lambda =316$ $nm$, broken lines), as the cylinder scans at constant
distance $d$ above the surface. The incidence is done with a $p$--polarized
Gaussian beam of $W=4000$ $nm$ at either $\theta _{0}=66^{o}$ (thick curves)
or $\theta _{0}=72^{o}$ (thin curves). The incident beam power (per unit
length) is $3.9313\times 10^{-6}$ $mW/nm$ on resonance and $3.9327\times
10^{-6}$ $mW/nm$ at $\lambda =316$ $nm$ when $\theta _{0}=66^{o}$, and
$3.9290\times 10^{-6}$ $mW/nm$ on resonance and $3.9315\times 10^{-6}$ $mW/nm$
at $\lambda =316$ $nm$ when $\theta _{0}=72^{o}$. As before, resonant
conditions provide a better image of the surface topography making the two
protrusions distinguishable with a contrast higher than the one obtained
without plasmon excitation. The surface image corresponding to the force
distribution is better when the protrusions (not shown here)
are inverted because then the
particle can be kept closer to the interface. Again, the curve contrast
yielded by protrusions and grooves is inverted from each other. The
positions of the force distribution peaks corresponding to the protrusions
now appear appreciably shifted with respect to the actual protrusions'
position. This shift is explained as due to the Goos--H\"{a}nchen effect of
the evanescent wave~\cite{yo01b}. We observe that the distance between these
peaks in the $F_{z}$ curve is aproximately $2X_{0}$. This shift is more
noticeable in the force distribution as the probe size
increases\footnote{For a better picture of this shift, see the
grating case in Ref.~\cite{yo01b}.}.
Again, the $F_{z}$ force distribution has a higher contrast at
the (shifted) position of the protrusions. The force signal with these
bigger particles is larger, but the probe has to be placed farther from the
surface at constant height scanning. This affects the strength of the
signal. Finally, it is important to state that the angle of incidence
(supposed to be larger than the critical angle $\theta _{c}$) influences
both the contrast and the strength of the force: the contrast decreases as
the angle of incidence increases. At the same time, the strength of the
force signal also diminishes.

As seen in the force figures for both sizes of particles, most curves
contain tiny ripples. They are due to the field intensity interference
pattern as shown in Figures~\ref{fig:prb3} and~\ref{fig:prb8}, and discussed
above. As the particle moves, the force on it is affected by this interference.
As a matter of fact, it can be noted in the force curves that these tiny
ripples are mainly present at the left side of the particle, which is the
region where stronger interference takes place.

It is worth remarking, however, that these oscillations are less marked in
the force distribution ({\it cf}. their tiny ripples), than in the near
field intensity distribution, where the interference patterns present much
higher contrast.

As stated in the previous section, evanescent fields and multiple scattering
are fruitful to extracting information from a detection
setup. The latter is, at the same time, somewhat
troublesome as it cannot be neglected
at will. This incovenient is well--known in NSOM, but it is diminished in PFM,
as remarked before.
The smoother signal provided by the force is underlined 
by two facts: one is the
averaging process on the particle surface, quantitatively interpreted from
the field surface integration involved in Maxwell's stress tensor. Other
is the local character of the force acting at each point of the particle
surface. Metallic particles are better candidates as probes of PFM
in comparison to dielectric particles,
since the force signal is not only enhanced at
resonance conditions, but it is also bigger and
presents better resolution. However,
dielectric particles are preferred when the distance to the interface is
large, since then the weak evanescent field present at these distances presents 
better coupling to the whispering--gallery modes than to the
plasmon surface waves of metallic particles.

\subsection{On the Attractive and Repulsive Nature of \newline
Vertical Forces and Their Orders of Magnitude}

\label{sec:discussion}

The horizontal forces acting on the particle are scattering forces due to
radiation pressure of both the incident evanescent wave and the field
scattered by the protrusions, thus the forces are positive in all the cases
studied. As for the vertical forces, two effects compete in determining
their sign. First, is the influence of the
polarizability~\cite{chaumet00b,chaumet00a},
which depends on the polarization of the
illumination. On the other hand, it is well known that an evanescent wave
produces only gradient forces in the vertical direction. For silver
cylinders, the force at wavelength $\lambda =387$ ($\epsilon
_{2}=-3.22+i0.70 $) and at $\lambda =441$ $nm$ ($\epsilon _{2}=-5.65+i0.75$)
must be attractive, while at $\lambda =316$ $nm$ ($\epsilon _{2}=0.78+i1.07$),
the real part of the polarizability changes its sign, and so does the
gradient force, thus becoming repulsive (on cylinders of not very large
sizes, as here). However, in the cases studied here, not only the
multiple scattering of light between the cylinder and the flat portion of
the interface, but also the surface defects, produce scattered waves both
propagating (into $z>0$) and evanescent under TIR conditions. Thus,
the scattering forces also contribute to the $z$--component of the force.
This affects the sign of the forces, but it is more significant as the size
of the objects increases. In larger cylinders and defects ({\it cf}.
Figure~\ref{fig:prb9}),
the gradient force is weaker than the scattering force thus
making $F_{z}$ to become repulsive on scanning at $\lambda =441$ $nm$
(plasmon excited). On the other hand, for the smaller silver cylinders
studied ({\it cf}. Figure~\ref{fig:prb5}), the gradient force is greater
than the scattering force at $\lambda =387$ $nm$ (plasmon excited), and thus
the force is attractive in this scanning. Also, as the distance between the
particle and the surface decreases, the gradient force becomes more
attractive~\cite{chaumet00b,chaumet00a}. This explains the dips and change
of contrast in the vertical force distribution on scanning both protrusions
and grooves. At $\lambda =316$ $nm$ (no plasmon excited), both scattering
and gradient forces act cooperatively in the vertical direction making the
force repulsive, no matter the size of the cylinder. For the silicon
cylinder, as shown, the vertical forces acting under TIR conditions are
attractive in absence of surface interaction (for both polarizations and the
wavelengths used). However, this interaction is able to turn into repulsive
the vertical force for $S$ polarization at $\lambda =538$ $nm$, due to the
scattering force.

This study also reveals the dependence of the attractive or repulsive nature
of the forces on the size of the objects (probe and defects of the
surface), apart from the polarizability of the probe and the distance to the
interface, when illumination under total internal reflection is considered.
The competition between the strength of the scattering and the gradient
force determines this nature.

The order of magnitude of the forces obtained in the preceding 2D
calculations is consistent with that of forces in experiments and 3D
calculations of Refs. \cite
{pohl93,guntherodt95,depasse92,sugiura93,kawata92,dereux94,girard94,almaas95,novotny97,hecht96,okamoto99,chaumet00b} and 
\cite{chaumet00a}.
Suppose a truncated cylinder with axial length $L=10$ $\mu m$, and a
Gaussian beam with $2W\sim 10$ $\mu m$. Then, a rectangular section of
$L\times 2W=10^{2}$ $\mu m^{2}$ is illuminated on the interface. For an
incident power $P_{0}\sim 1$ $mW$, spread over this rectangular section, the
incident intensity is $I_{0}\sim 10^{-2}$ $mW/\mu m^{2}$, and the force
range from our calculations is $F\sim 10^{-2}-10^{-1}$ $pN$. Thus, the
forces obtained in Figures~\ref{fig:prb9}(b) and~\ref{fig:prb9}(d) are
consistent with those presented, for example, in Ref.~\cite{kawata92}.

\section{Concluding Remarks \& Future Prospects}
\label{sec:concluding}

The forces exerted by both propagating and evanescent fields on small particles
are the basis to understand estructural characteristics of time--harmonic
fields. The simplest evanescent field that can be built
is the one that we have illustrated on transmission at a dielectric
interface when TIR conditions occur.
Both dielectric and metallic particles are pushed along the direction of
propagation of the evanescent field, independently of the size
(scattering and absorption forces).
By contrast, forces behave differently along
the decay direction (gradient forces)
on either dielectric and metallic particles, as studied for
dipolar--sized particles, Section~\ref{sec:dipapprox}. The analysis done
in presence of a
rough interface, and with particles able to interact with it show that
scattering, absorption and gradient forces act both in the amplitude
and phase directions, when multiple scattering takes place.
Moreover, the excitation of particle
resonances enhances this interaction, and, at the
same time, generates evanescent fields (surface waves)
on their surface, which makes even
more complex this mixing among force components. Thus, an analysis based on a
small particle isolated is not feasible due to the high
inhomogeneity of the field.

It is, however, this inhomogeneity what provides a way to imaging a surface
with structural features, such as topography.
The possibility offered by the combination of evanescent fields and Mie
resonances is however
not unique. As inhomgeneous fields (which can be analitically
decomposed into propagating and evanescent
fields~\cite{nieto-vesperinas91}) play
an important role in the mechanical action of the electromagnetic wave
on dielectric particles (either on or out of resonance),
they can be used to operate at the nanometric scale on such entities,
to assist the formation of ordered particle structures as for
example~\cite{burns89,burns90,antonoyiannakis97,antonoyiannakis99,malley98,bayer98,barnes02},
with help of
these resonances. Forces created by evanescent fields on particles
and morphology--dependent resonances are the keys to control the optical
binding and the formation of photonic molecules. Also, when a particle is
used as a nanodetector, these forces are the signal in a scheme of photonic
force microscopy as modeled in this article. It has been shown that the 
evanescent
field forces and plasmon resonance excitations permit to manipulate
metallic particles~\cite{novotny97,chaumet01,chaumet02}, as well as to
make such microscopy~\cite{yo01c}. Nevertheless, controlled experiments
on force magnitudes, both due to evanescent and propagating waves,
are yet scarce
and thus desirable to be fostered.

The concepts released
in this article open an ample window to investigate on soft matter components
like in cells and molecules in biology.
Most  folding processes require small forces to detect and control
in order not to alter them and be capable of actuating and extracting
information from them.

\section*{General Annotated References}
\label{sec:references}

Complementary information and sources for some of the contents treated in this
report can be found in next bibliography:

\begin{itemize}
\item Electromagnetic optics: there are many books where to find the basis of
      the electromagnetic theory and optics. We cite here the most
      common:~\cite{jackson75,bornwolf99}. The Maxwell's Stress
      Tensor is analysed in~\cite{jackson75,stratton41}. The mathematical level
      of these textbooks is similar to the one in this report.
\item Mie theory can be found in~\cite{vandehulst81,kerker69,bohren83}.
      In these
      textbooks, Optics of particles is develop with little mathematics and
      all of them are comparable in contents.
\item Resonances can be understood from the textbooks before, but a more
      detailed information, with applications and the implications in many
      topics can be found in the following
      references:  \cite{yophd,hill88,barber90}.
      The last two references are
      preferently centred in dielectric particles. The first one compiles
      some of the information in these two references and some other from
      scientific papers. Surface plasmons, on the
      other hand, are studied in depth in~\cite{raether88}. They are
      easy to understand from general physics.
\item Integral equations in scattering theory and angular spectrum
      representation (for the decomposition of time--harmonic fields in
      propagating and evanescent components)
      are treated in~\cite{nieto-vesperinas91}.
      The mathematical level is similar to the one in this report.
\item The Coupled Dipole Method can be found in the scientific papers cited
      in Section~\ref{sec:CDM}. A More didactical reference
      is~\cite{chaumetphd,rahmaniphd}.
      The information shed in this report on the CDM is extended in
      these references.
\item The dipolar approximation, in the context of optical forces and
      evanescent fields, can be complemented in scientific
      papers:~\cite{gordon73,chaumet00c,yo02b} and in the
      monograph:~\cite{novotny00}.
\item A more detailed discussion on the sign of optical forces for dipolar
      particles, as well as larger elongated particles, can be found in the
      scientific papers~\cite{yo02a,yo02b,chaumet00b}.
\item Monographs on NSOM and tweezers:~\cite{nieto-vesperinas96,sheetz97}.
\end{itemize}

\vspace{2cm} {\Large {\bf Acknowledgments}}
\newline
\newline
We thank P. C. Chaumet and M. Lester for work that we have shared through
the years. Grants from DGICYT and European Union, as well as a fellowship of
J. R. Arias-Gonz\'alez from Comunidad de Madrid, are also acknowledged.

\end{document}